\documentclass[aps,prc,amsmath,amssymb,superscriptaddress,twocolumn,nofootinbib,showpacs,showkeys,preprintnumbers,notitlepage]{revtex4-2}
\usepackage[usenames,dvipsnames,svgnames,table]{xcolor}
\usepackage{graphicx}
\usepackage{dcolumn}
\usepackage{bm}
\usepackage{hyperref}
\hypersetup{ 
    pdfnewwindow=true,      
    colorlinks=true,       
    allcolors=[RGB]{31 119 180}
} 
\usepackage[capitalise]{cleveref}
\usepackage{dsfont}
\usepackage{placeins}
\usepackage{todonotes}
\usepackage[utf8]{inputenc}
\usepackage{gensymb}
\usepackage{multirow}
\usepackage{amsmath}
\usepackage{amsfonts}
\usepackage{amssymb}
\usepackage{enumitem,amssymb}
\usepackage{array}   
\newcolumntype{L}{>{$}l<{$}} 
\newcolumntype{R}{>{$}r<{$}}
\newcolumntype{C}{>{$}c<{$}}
\usepackage{xspace}
\usepackage{braket}
\usepackage{units}
\usepackage{slashed}
\usepackage[normalem]{ulem}
\usepackage[format=plain,justification=RaggedRight,singlelinecheck=false]{caption}
\usepackage[format=plain,justification=centering,singlelinecheck=false]{subcaption}
\usepackage{tabularx}
\usepackage{nameref}

\usepackage{xpatch}
\makeatletter
\xpatchcmd{\@ssect@ltx}{\@xsect}{\protected@edef\@currentlabelname{#8}\@xsect}{}{}
\xpatchcmd{\@sect@ltx}{\@xsect}{\protected@edef\@currentlabelname{#8}\@xsect}{}{}
\makeatother
\usepackage{hyperref}

\graphicspath{{figs/}}

\newcommand{\mevnospace}{\ensuremath{{\mathrm{\,Me\kern -0.1em V}}}}
\newcommand{\gevnospace}{\ensuremath{{\mathrm{\,Ge\kern -0.1em V}}}}
\newcommand{\tevnospace}{\ensuremath{{\mathrm{\,Te\kern -0.1em V}}}}

\newcommand\bsub{\begin{subequations}}
\newcommand\esub{\end{subequations}}

\newlist{todolist}{itemize}{2}
\setlist[todolist]{label=$\square$}
\usepackage{pifont}

\usepackage{orcidlink}

\usepackage{caption}
\usepackage{subcaption}

\usepackage{tikz}
\usepackage[compat=1.1.0]{tikz-feynman}

\newcommand{\ub}{Departament de F\'isica Qu\`antica i Astrof\'isica and Institut de Ci\`encies del Cosmos (ICCUB), Facultat de F\'isica,  Universitat de Barcelona, 
Barcelona, Spain}
\newcommand{\jlab}{Theory Center, Thomas Jefferson National Accelerator Facility, 
Newport News, VA 
, USA}
\newcommand{\ice}{Institute of Space Sciences (ICE, CSIC), 
Barcelona, Spain}
\newcommand{\ieec}{Institut d'Estudis Espacials de Catalunya (IEEC),
Barcelona, Spain}
\newcommand{\fias}{Frankfurt Institute for Advanced Studies,
Frankfurt am Main, Germany
}

\begin{document}

\preprint{JLAB-THY-23-3874}

\title{Recent progress on in-medium properties of heavy mesons \\ from finite-temperature EFTs} 

\author{Gl\`oria Monta\~na}
\email{gmontana@jlab.org}
\affiliation{\ub}
\affiliation{\jlab}
\author{\`Angels Ramos}
\affiliation{\ub}
\author{Laura Tolos}
\affiliation{\ice}
\affiliation{\ieec}
\affiliation{\fias}
\author{Juan~M. Torres-Rincon}
\affiliation{\ub}

\begin{abstract}
Mesons with heavy flavor content are an exceptional probe of the hot QCD medium produced in heavy-ion collisions. In the past few years, significant progress has been made toward describing the modification of the properties of heavy mesons in the hadronic phase at finite temperature. Ground-state and excited-state thermal spectral properties can be computed within a self-consistent many-body approach that employs appropriate hadron-hadron effective interactions, providing a unique opportunity to confront hadronic Effective Field Theory predictions with recent and forthcoming lattice QCD simulations and experimental data.
In this article, we revisit the application of the imaginary-time formalism to extend the calculation of unitarized scattering amplitudes from the vacuum to finite temperature. These methods allow us to obtain the ground-state thermal spectral functions. The thermal properties of the excited states that are dynamically generated within the molecular picture are also directly accessible. We present here the results of this approach for the open-charm and open-bottom sectors. 
We also analyze how the heavy-flavor transport properties, which are strongly correlated to experimental observables in heavy-ion collisions, are modified in hot matter. In particular, transport coefficients can be computed using an off-shell kinetic theory that is fully consistent with the effective theory describing the scattering processes. The results of this procedure for both charm and bottom transport coefficients are briefly discussed. 
\end{abstract}


\maketitle

\section{Introduction}\label{sec:intro}


The discovery in 2003 of the charm-strange mesons $D_{s0}^*(2317)$~\cite{BaBar:2003oey} and $D_{s1}(2460)$\cite{CLEO:2003ggt}, with masses significantly lower than the quark-model predictions for the lowest lying scalar and axial-vector $c\bar{s}$ mesons, has generated intensive discussions on their internal structure for the past twenty years. Together with the $X(3872)$ charmonium-like state, which was first observed also in 2003~\cite{Belle:2003nnu}, they are the first candidates of exotic mesons with multiquark content in the heavy meson sector.
Despite the enormous efforts, there exists still a lack of consensus on whether the $D_{s0}^*(2317)$ and $D_{s1}(2460)$ are meson molecules, compact tetraquarks or an admixture with $c\bar{s}$ components. Yet there are compelling arguments in favor of the molecular interpretation: 
their masses lie very close to the $DK$ and $D^*K$ thresholds, respectively, and the mass difference between these two excited states is very similar to that between the $D$ and $D^*$ ground states ($\sim 140$~MeV). Therefore the prevailing picture is that they have a large component of molecular $DK$ or $D^*K$ and coupled channels~\cite{Barnes:2003dj,Szczepaniak:2003vy,Kolomeitsev:2003ac,Hofmann:2003je,Guo:2006fu,Gamermann:2006nm,Faessler:2007gv,Flynn:2007ki}, which is supported by lattice QCD data~\cite{Mohler:2013rwa,Lang:2014yfa,Bali:2017pdv,Cheung:2020mql}.

Closely related is the case of the broad structures observed in the $D\pi$ and $D^*\pi$ invariant mass distributions~\cite{Belle:2003nsh,FOCUS:2003gru,BaBar:2009pnd,LHCb:2015klp} and reported as the $D_0^*(2300)$ and $D_1(2430)$ states by 
the Particle Data Group (PDG)~\cite{pdg}. The value reported for the mass of the $D_0^*(2300)$ strongly depends on the production mechanism, ranging from $\sim 2400$~MeV with $\gamma\,A$ reactions to $\sim 2300$~MeV from $B$-meson decays, and the values reported by the LHCb collaboration for the charged partner lie in the middle. 
The fact that these values are close to the mass of the $D_{s0}^*(2317)$, or are even larger, is in contradiction with constituent quark model predictions. An answer to this puzzle is naturally provided by the use of unitarized effective models in coupled channels, which give rise to two $D_0^*$ poles in the energy-region of the $D_{0}^*(2300)$, and two $D_1$ poles in that of the $D_1(2430)$~\cite{Kolomeitsev:2003ac,Guo:2006fu,Guo:2009ct,Albaladejo:2016lbb,Du:2020pui,Asokan:2022usm}. Strong evidence that the $D_0^*(2300)$ and $D_1(2430)$ states could be interpreted as meson molecules with a two-pole structure comes from the remarkably good agreement that the authors of Ref.~\cite{Albaladejo:2016lbb} found with the lattice QCD results of the lowest-lying energy levels of Ref.~\cite{Moir:2016srx}. More recently the authors of Ref.~\cite{Asokan:2022usm} showed that, in addition to the pole reported in~\cite{Moir:2016srx}, a second pole on an unphysical Riemann sheet is needed in the analysis of the lattice data.

While heavy-quark spin symmetry (HQSS) is responsible for the near degenerate patterns between the open-charm scalars, $D_{s0}^*(2317)$ and $D_0^*(2300)$, and between the axial vectors, $D_{s1}(2460)$ and $D_1(2430)$, from heavy-quark flavor symmetry (HQFS) one expects to find a similar degeneracy in the bottom sector. For instance, unitarized effective field theory (EFT) models that find the $D_{s0}^*(2317)$ as a $DK$ bound state predict a bottom partner, a $\bar BK$ bound state, with a similar binding energy~\cite{Kolomeitsev:2003ac,Altenbuchinger:2013vwa,Torres-Rincon:2014ffa}, in agreement with lattice QCD results~\cite{Lang:2015hza}.
The bottomed analogues of the scalar $D_{s0}^*(2317)$ and $D_0^*(2300)$ are still to be found experimentally, but the axial-vector $B_1(5721)$ and $B_{s1}(5830)$ could presumably be the bottom-flavor partners of the $D_1(2430)$ and $D_{s1}(2460)$.

A new venue to study the nature of heavy-flavor exotica has recently emerged with relativistic heavy-ion collisions (HICs), where an extremely hot quark-gluon plasma (QGP) is created. At high collision energies, such as those at the Relativistic Heavy-Ion Collider (RHIC) and the Large Hadron Collider (LHC), abundant heavy quark-antiquark pairs are produced in the initial hard scattering between partons. These pairs then propagate through the rapidly expanding and cooling QGP. At a temperature of about $T_c = 156$~MeV the hadronic medium is eventually formed, and the interactions between the heavy hadrons and the surrounding light mesons occur until the so-called kinetic freeze-out at lower temperatures. This offers an excellent opportunity to test the in-medium properties of the heavy mesons produced, including those of heavy exotica. Furthermore, the novel employment of femtoscopy techniques in $pp$, $pA$ and $AA$ collisions at the LHC and RHIC to determine the scattering parameters of $D$ mesons with light-flavor hadrons will certainly help probe the hadronic interactions, as well as the effects of the hadronic medium  \cite{STAR:2014dcy,STAR:2018uho,ALICE:2018ysd,ALICE:2019eol,ALICE:2019gcn,ALICE:2020mfd,Fabbietti:2020bfg,Battistini2023}.

The non-perturbative regime of hot hadronic matter can be consistently treated using effective Lagrangians combined with quantum field theory techniques at finite temperature, often denominated as thermal EFTs. While finite-temperature lattice QCD has been for many years a powerful theoretical source of information on hot QCD matter, thermal EFTs are a complementary tool that enable us to approach the QCD phase transition from the chirally-broken phase of hadrons. 

In this work we use thermal EFTs to access the finite-temperature properties of charm and bottom mesons in a hot medium. To this end, we will revisit the calculations that we presented in a series of works on charmed mesons~\cite{Montana:2020lfi,Montana:2020vjg,Torres-Rincon:2021yga,MontanaFaiget:2022cog} and, in addition, we will present extensions of the calculations to the heavy mesons with bottom flavor~\cite{MontanaFaiget:2022cog}. We note that, although it will not be discussed here, our findings in the charm sector have been checked against lattice QCD calculations at the level of Euclidean correlators~\cite{Montana:2020var} and that we have also studied the thermal modification of the $X(3872)$ exotic state and its spin-flavor partners when these are assumed to be of molecular nature~\cite{Montana:2022inz}.

The rest of the article is organized as follows. In Section~\ref{sec:formalism} we discuss the main ingredients of our thermal EFT approach to address the in-medium properties of open heavy-flavor mesons: The effective hadron-hadron interactions, the use of the imaginary-time formalism to evaluate properties of a system in thermal equilibrium (e.g. thermal corrections to the mass and decay width), and the evolution in real-time to tackle the description of a system out of equilibrium and compute transport coefficients. We will review the key ideas and refer to our previous works for technical details~\cite{Montana:2020lfi,Montana:2020vjg,Torres-Rincon:2021yga}. Section~\ref{sec:results} presents novel results for bottomed mesons. These include self-energies, spectral functions and transport coefficients, alongside a comparison with selected results from our previous works in the charm sector. We end with a final discussion and conclusions in Section~\ref{sec:conclusions}. 

\section{Formalism}\label{sec:formalism}

In this work we use a thermal effective field theory approach that was developed in a series of works~\cite{Montana:2020lfi,Montana:2020vjg} for charmed hadrons. It is based on unitarized heavy-meson chiral perturbation theory (HMChPT) combined with thermal field theory techniques using the imaginary-time formalism to address the thermal effects on the properties of heavy mesons in a mesonic medium at finite temperature. The kinetic theory describing the heavy-meson dynamics in the hot medium can be derived using the real-time formalism~\cite{Torres-Rincon:2021yga}. The resulting kinetic equation depends on thermal scattering amplitudes and spectral functions. For the calculation of transport coefficients it is sufficient to assume a system near equilibrium and employ equilibrium quantities. 
In the following we summarize the main steps to compute some relevant quantities at finite temperature.

\subsection{Interactions between open heavy-flavor mesons and light mesons}\label{sec:interactions}

We start by outlining the main features of the interactions between open-heavy flavor mesons, i.e. mesons with one charm or bottom quark $H\ni\{D,D_s,D^*,D_s^*,\bar B,\bar B_s,\bar B^*,\bar B_s^*\}$, and the light Goldstone bosons $\Phi\ni \{\pi,K,\bar K,\eta\}$ within the framework of HMChPT, an effective field theory that has been widely used in the last years to describe the interactions between open-heavy flavor mesons and light mesons \cite{Wise:1992hn,Burdman:1992gh,Casalbuoni:1996pg,Kolomeitsev:2003ac,Lutz:2007sk}, with chiral symmetry and heavy-quark spin-flavor symmetry (HQSFS) as its guiding principles. When combined with a unitarization technique such as the solution of the Bethe-Salpeter equation in the on-shell factorization scheme~\cite{Oller:1997ti,Oset:1997it}, the potentials of the HMChPT lead to the dynamical generation of quasi-bound states from the $s$-wave scattering of heavy flavored mesons off Goldstone bosons. In particular, it provides a description of the lightest scalar and axial vector open-charm states (i.e. $D_0^*(2300)$, $D_{s0}(2317)$, $D_1(2430)$ and $D_{s1}(2460)$) as hadronic molecules, as well as predictions for their counterparts in the open-bottom sector. 

The Lagrangian of HMChPT expanded at next-to-leading order (NLO) in the chiral expansion and at leading order (LO) in the inverse of the mass of the heavy meson $m_H$ reads \cite{Guo:2009ct,Geng:2010vw,Abreu:2011ic,Liu:2012zya,Tolos:2013kva},
\begin{equation}
    \mathcal{L} =\mathcal{L}_\textrm{LO}+\mathcal{L}_\textrm{NLO} \ ,
\end{equation}
with the subscripts LO and NLO referring to the chiral power counting, and
\begin{widetext}
\begin{subequations}
\begin{align}\label{eq:LagrangianLO}
\mathcal{L}_\textrm{LO}&\ =\mathcal{L}^{\textrm{ChPT}}_\textrm{LO}+\langle\nabla^\mu H\nabla_\mu H^\dagger\rangle-m_H^2\langle HH^\dagger\rangle-\langle\nabla^\mu H^{*\nu}\nabla_\mu H^{*\dagger}_{\nu}\rangle+m_H^2\langle H^{*\nu}H^{*\dagger}_{\nu}\rangle \nonumber \\
 &\quad +ig\langle H^{*\mu}u_\mu H^\dagger-Hu^\mu H^{*\dagger}_\mu\rangle+\frac{g}{2m_D}\langle V^*_\mu u_\alpha\nabla_\beta H^{*\dagger}_\nu-\nabla_\beta V^*_\mu u_\alpha H^{*\dagger}_\nu\rangle\epsilon^{\mu\nu\alpha\beta} \ , \\
\label{eq:lagrangianNLO}\nonumber
 \mathcal{L}_\textrm{NLO}&\ = \mathcal{L}^{\textrm{ChPT}}_\textrm{NLO} -h_0\langle HH^\dagger\rangle\langle\chi_+\rangle+h_1\langle H\chi_+H^\dagger\rangle+h_2\langle HH^\dagger\rangle\langle u^\mu u_\mu\rangle  +h_3\langle Hu^\mu u_\mu H^\dagger\rangle \\ \nonumber
 &\quad +h_4\langle\nabla_\mu H\nabla_\nu H^\dagger\rangle\langle u^\mu u^\nu\rangle+h_5\langle\nabla_\mu H\{u^\mu,u^\nu\}\nabla_\nu H^\dagger \rangle \\ \nonumber
 &\quad +\tilde{h}_0\langle H^{*\mu}H^{*\dagger}_\mu\rangle\langle\chi_+\rangle-\tilde{h}_1\langle H^{*\mu}\chi_+H^{*\dagger}_\mu\rangle-\tilde{h}_2\langle H^{*\mu}H^{*\dagger}_\mu\rangle\langle u^\nu u_\nu\rangle -\tilde{h}_3\langle H^{*\mu}u^\nu u_\nu H^{*\dagger}_\mu\rangle  \\ 
 &\quad -\tilde{h}_4\langle\nabla_\mu H^{*\alpha}\nabla_\nu H^{*\dagger}_\alpha\rangle\langle u^\mu u^\nu\rangle-\tilde{h}_5\langle\nabla_\mu H^{*\alpha}\{u^\mu,u^\nu\}\nabla_\nu H^{*\dagger}_\alpha\rangle,
\end{align}
\end{subequations}
\end{widetext}
where $\mathcal{L}^{\textrm{ChPT}}_\textrm{LO}$ and $\mathcal{L}^{\textrm{ChPT}}_\textrm{NLO}$ encode the chiral Lagrangians of the pure light-meson sector. In the charm sector, $H$ and $H^*_\mu$ denote the antitriplets of pseudoscalar $D$-mesons, $\left(D^0 \; D^+ \; D^+_s\right)$, and vector $D^*$-mesons, $\left(D^{*0}_\mu \; D^{*+}_\mu \; D^{*+}_{s,\mu}\right)$, respectively, while in the bottom sector they correspond to the pseudoscalar $\bar{B}$-mesons, $\left(\bar B^- \; \bar B^0 \; \bar B^0_s\right)$, and vector $\bar B^*$-mesons, $\left(\bar B^{*-}_\mu \; \bar B^{*0}_\mu \; \bar B^{*0}_{s,\mu}\right)$. The octet of Goldstone bosons are contained in the unitary matrix $u=\exp (i\Phi/\sqrt{2} f_\pi)$ in the building blocks $u_\mu=i(u^\dagger\partial_\mu u-u\partial_\mu u^\dagger)$ and $\chi_+=u^\dagger\chi u^\dagger+u\chi u$, with the quark mass matrix $\chi=\textrm{diag}(m_\pi^2,m_\pi^2,2m_K^2-m_\pi^2)$.
For our calculations, we rely on the values from the Fit-2B in Ref.~\cite{Guo:2018tjx}. We employ the relation $\{h_i\}=\{\tilde{h}_i\}$ that is applicable at LO in the heavy-quark mass expansion For the specific values, please refer to our previous works~\cite{Montana:2020lfi,Montana:2020vjg}. In the bottom sector we take advantage of the heavy-quark mass scaling of the low energy constants (LECs), $\{h_i^B\}\hat{M}_B^{-1}= \{h_i^D\}\hat{M}_D^{-1}$, for $h^H_i\in\{h^H_0,h^H_2,h^H_3,h^H_4\hat{M}_H^2,h^H_5\hat{M}_H^2\}$.

The tree-level potential for the process $H^{(*)}\Phi\to H^{(*)}\Phi$ reads
\begin{align} \nonumber\label{eq:potential}
 \mathcal{V}^{ij}&(s,t) = \frac{1}{f_\pi^2}\Big[\frac{C_\textrm{LO}^{ij}}{4}(s-u)-4C_0^{ij}h_0+2C_1^{ij}h_1\nonumber \\
 &\ -2C_{24}^{ij}\Big(2h_2(p_2\cdot p_4) +h_4\big((p_1\cdot p_2)(p_3\cdot p_4)\nonumber \\ &\ +(p_1\cdot p_4)(p_2\cdot p_3)\big)\Big) +2C_{35}^{ij}\Big(h_3(p_2\cdot p_4)\nonumber \\ &\ +h_5\big((p_1\cdot p_2)(p_3\cdot p_4)+(p_1\cdot p_4)(p_2\cdot p_3)\big)\Big)
 \Big],
\end{align}
where $s=(p_1+p_2)^2$, $t=(p_1-p_3)^2$, and $u=(p_1-p_4)^2$ are the standard Mandelstam variables, and the superindices $i,j$ denote the incoming and outgoing channels from the coupled-channel basis. For instance, in the sector with strangeness $S=0$ and isospin $I=1/2$, which are the quantum numbers of the $D_0^*(2300)$, we have $\{D\pi,D\eta,D_s\bar{K}\}$, and for $S=1$ and $I=0$, as for the $D_{s0}(2317)$, we have $\{DK,D_s\eta\}$. 
We refer the reader to Ref.~\cite{Montana:2020vjg} for the values of the coefficients $C^{ij}_k$ in the isospin basis. 

The partial-wave projection with angular momentum $\ell$ is then obtained through the relation
\begin{equation}
  \mathcal{V}^{ij}_{\ell}(s)= \frac{1}{2} \int_{-1}^{+1} d(\cos\theta) \ P_{\ell} (\cos \theta) \mathcal{V}^{ij}(s,t(s,\cos \theta))  \ , \label{eq:proj}
\end{equation}
where $\theta$ is the scattering angle between the initial and final particles in the center of mass, and $P_\ell(\cos\theta)$ the Legendre polynomial of order $\ell$.

\subsection{Thermal equilibrium properties}\label{sec:thermalequilibrium}

A hadron gas forms once the temperature is turned on. The relative abundance of each hadron species in thermal equilibrium is determined by the corresponding thermal distribution functions, i.e. the Bose-Einstein distribution for mesons and the Fermi-Dirac distribution for baryons. At temperatures $T\sim 100-150$~MeV light mesons become the primary components of the medium and heavy mesons behave as Brownian particles scattering of the light mesons. We exploit this scenario and employ the HMChPT potential presented above to describe the dynamics of the heavy mesons with the light mesons in the bath.

In order to incorporate the effects of the hot medium, it is necessary to follow the techniques of thermal field theory. There exist two complementary formulations of thermal field theory that can be used to describe a system in thermal equilibrium, the ``imaginary-time'' and the ``real-time'' formalisms. In the imaginary-time formalism (ITF), also called Matsubara formalism owing to the pioneering work by Matsubara~\cite{Matsubara:1955ws}, time is treated as a purely imaginary quantity and then one performs an analytical continuation from Euclidean to Minkowski spacetime at the end of the calculation. In the real-time formalism, in contrast, the calculation is done in Minkowski spacetime, considering explicitly the evolution in real time~\cite{kadanoff1962quantum,Keldysh:1964ud}. While the latter is capable of describing systems even outside thermal equilibrium, and thus appropriate to address out-of-equilibrium properties and transport coefficients, as we will discuss in the next section, the ITF has the advantage of resembling in a more intuitive way the zero temperature field theory. For instance, the main difference in the form of the propagators and the diagrammatic structure of the perturbative expansion, in the ITF, is the acquisition of thermal weights in the phase-space integrals compared with those at $T=0$, as we will show below.
An approach based on the ITF has been developed in the recent years to study the properties of open heavy-flavor mesons in hot hadronic matter at vanishing baryonic density~\cite{Montana:2020lfi,Montana:2020vjg,Cleven:2017fun}.

To compute the thermal corrections to a given quantity, such as the two-meson propagator or the self-energy,
the ITF provides some simple rules that basically consist in replacing the zeroth component of the four-momenta of the particles by discrete Matsubara frequencies $i\omega_n$, with $\omega_n=2n\pi/\beta$ for bosons and $\beta=1/T$, and transforming the integration over internal energies into a summation over Matsubara frequencies. Then, by using some established computational techniques based on contour integrals and analytic continuation, the calculations can be done similarly as in the vacuum field theory.
For details, we encourage the reader to consult the classical Refs.~\cite{Weldon:1983jn,Das:1997gg,kapustagale,lebellac,fetterwalecka}.

Using the rules above, the thermal two-meson propagator takes the form
\begin{widetext}
    \begin{align}\label{eq:loop}
 \mathcal{G}(E,\bm{p}\,;T)&= 
 \int\frac{d^3q}{(2\pi)^3}  \int 
 d\omega\int d\omega' \ \frac{\mathcal{S}_H(\omega,\bm{q}\,;T)\mathcal{S}_\Phi(\omega',\bm{p}-\bm{q}\,;T)}{E-\omega-\omega'+i\varepsilon} 
 \left[1+f(\omega,T)+f(\omega',T)\right] \ ,
\end{align}
\end{widetext}

with $p^\mu=(E,\bm{p})$ the momentum in the center of mass of the two-meson system.
In the above equation, in addition to the medium corrections arising from the ITF, i.e. the additional weighting factors containing appropriate combinations of Bose-Einstein distribution functions $f(\omega,T)=(e^{\beta\omega}-1)^{-1}$, the meson masses are dressed by the spectral functions $\mathcal{S}_H$ and $\mathcal{S}_\Phi$. Note that this is a compact expression with the integrals over energy extending from $-\infty$ to $+\infty$. 

In the case of zero temperature, it is customary to regularize the vacuum contribution to the two-meson propagator, for example, by introducing a hard cutoff in the three-momentum integration. We use $\Lambda=800$~MeV, a value that corresponds to the scale of the degrees of freedom that were integrated out when constructing the meson-meson interaction amplitude from the effective Lagrangian, and that is consistent with the regularization scheme used in Ref.~\cite{Guo:2018tjx}, from where we adopted the values for the LECs of the NLO potential. This regularization scheme is straightforward to extend to finite temperature.
The tree-level potential in Eq.~(\ref{eq:potential}) does not change at finite temperature because in the ITF thermal corrections enter in loop diagrams~\cite{kapustagale,lebellac}. 

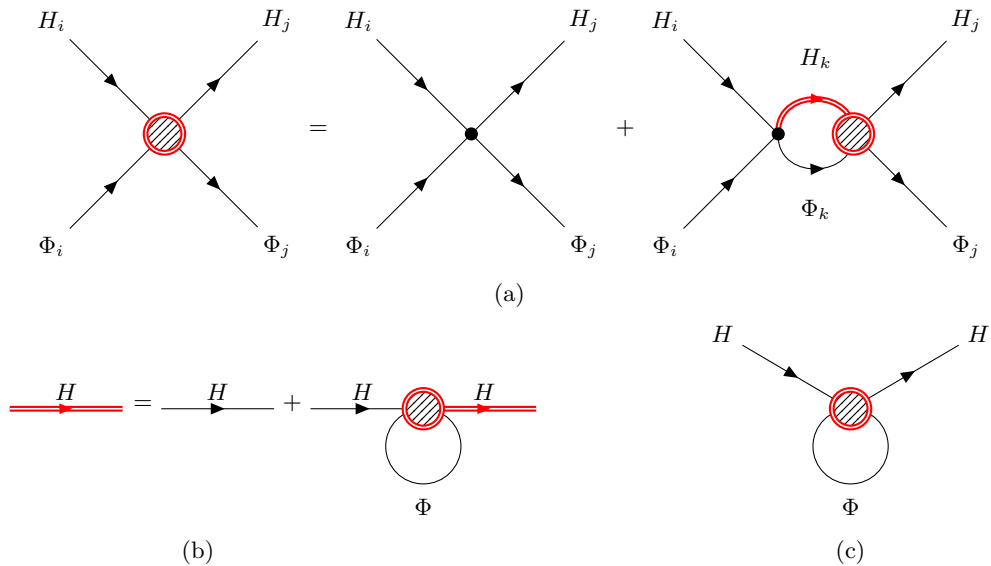
\begin{figure*}
\centering
\begin{subfigure}[b]{\textwidth}\centering
 \begin{tikzpicture}[baseline=(i.base)]
    \begin{feynman}[small]
      \vertex (i)  at (0, 0) {};
      \vertex (a) at (-1.5,1.5) {\(H_i\)};
      \vertex (b) at (1.5,1.5) {\(H_j\)};
      \vertex (d) at (1.5,-1.5) {\(\Phi_j\)};
      \vertex (c) at (-1.5,-1.5) {\(\Phi_i\)};
      \diagram*{
        (a) -- [fermion] (i), 
        (i) -- [fermion] (b),
        (c) -- [fermion] (i),
        (i) -- [fermion] (d),
       };
     \draw[blob, white] (i) circle(0.25);
     \draw[blob, thick,double,red,pattern= north east lines] (i) circle(0.25);
    \end{feynman}
  \end{tikzpicture}
  $=$
  \begin{tikzpicture}[baseline=(i.base)]
    \begin{feynman}[small]
      \vertex[dot] (i) at (0,0) {};
      \vertex (a) at (-1.5,1.5) {\(H_i\)};
      \vertex (b) at (1.5,1.5) {\(H_j\)};
      \vertex (d) at (1.5,-1.5) {\(\Phi_j\)};
      \vertex (c) at (-1.5,-1.5) {\(\Phi_i\)};
      \diagram*{
        (a) -- [fermion] (i), 
        (i) -- [fermion] (b),
        (c) -- [fermion] (i),
        (i) -- [fermion] (d),
       } ;    
     \draw[dot,black,fill=black] (i) circle(.8mm);
    \end{feynman}
  \end{tikzpicture}
  $+$
  \begin{tikzpicture}[baseline=(i.base)]
    \begin{feynman}[small]
      \vertex[dot] (i) at (0,0) {};
      \vertex (a) at (-1.5,1.5) {\(H_i\)};
      \vertex (j) at (1,0);
      \vertex (b) at (2.5,1.5) {\(H_j\)};
      \vertex (d) at (2.5,-1.5) {\(\Phi_j\)};
      \vertex (c) at (-1.5,-1.5) {\(\Phi_i\)};
      \vertex (b1) at (0.5,1) {\(H_k\)};
      \vertex (d1) at (0.5,-1) {\(\Phi_k\)};
      \diagram*{
        (a) -- [fermion] (i), 
        (j) -- [fermion] (b),
        (i) -- [with arrow=0.5,arrow size=0.12em,red, thick, double, half left, looseness=1.5] (j),
        (i) -- [fermion,half right, looseness=1.5] (j),
        (c) -- [fermion] (i),
        (j) -- [fermion] (d),
       } ;    
     \draw[dot,black,fill=black] (i) circle(0.8mm);
     \draw[blob, white] (j) circle(0.25);
     \draw[blob, thick,double,red,pattern= north east lines] (j) circle(0.25);
    \end{feynman}
  \end{tikzpicture}
\caption{}
\label{fig:diagrams-a}
\end{subfigure}
\begin{subfigure}[b]{0.5\textwidth}\centering\hfill
 \begin{tikzpicture}[baseline=(a.base)]
    \begin{feynman}[small]
      \vertex (a) at (0,0);
      \vertex (b) at (1.5,0);
      \diagram* {
        (a) -- [red, double, thick,with arrow=0.5,arrow size=0.12em,edge label=\(\textcolor{black}{H}\)] (b), 
       };
    \end{feynman}
  \end{tikzpicture}
  $=$
 \begin{tikzpicture}[baseline=(a.base)]
    \begin{feynman}[small]
      \vertex (a) at (0,0);
      \vertex (b) at (1.5,0);
      \diagram* {
        (a) -- [fermion, edge label=\(H\)] (b), 
       };
    \end{feynman}
  \end{tikzpicture}
  $+$
  \begin{tikzpicture}[baseline=(a.base)]
    \begin{feynman}[small, inline=(a)]
      \vertex (i)  at (0, 0) {};
      \vertex (a) at (-1.5,0);
      \vertex (b) at (1.5,0);
      \vertex (d) at (0,-0.5);
      \vertex (e) at (0,-1.3) {\(\Phi\)};
      \diagram* {
        (a) -- [fermion,edge label=\(H\)] (i), 
        (i) -- [red, double, thick,with arrow=0.5,arrow size=0.12em,edge label=\(\textcolor{black}{H}\)] (b),
       } ;   
     \draw (d) circle(0.5);
     \draw[blob, white] (i) circle(0.25);
     \draw[blob, thick,double,red,pattern= north east lines] (i) circle(0.25);
    \end{feynman}
  \end{tikzpicture}
\caption{\quad}
\label{fig:diagrams-b}
\end{subfigure}
\begin{subfigure}[b]{0.45\textwidth}\centering
  \begin{tikzpicture}
    \begin{feynman}[small]
      \vertex (i) at (0,0) {} ;
      \vertex (a) at (-1.7,1) {\(H\)};
      \vertex (b) at (1.7,1) {\(H\)};
      \vertex (d) at (0,-0.5);
      \vertex (e) at (0,-1.3) {\(\Phi\)};
      \diagram* {
        (a) -- [fermion] (i), 
        (i) -- [fermion] (b),
       } ;   
     \draw (d) circle(0.5);
     \draw[blob, white] (i) circle(0.25);
     \draw[blob, thick,double,red,pattern= north east lines] (i) circle(0.25);
    \end{feynman}
  \end{tikzpicture}
\caption{}
\label{fig:diagrams-c}
\end{subfigure}\hfill
\caption{\textbf{1a.} Diagrammatic representation of the Bethe-Salpeter equation in coupled channels in Eq.~(\ref{eq:bethesalpeter}). At finite temperature, the $\mathcal{T}$ matrix (hatched blob) is obtained from the unitarization of the interaction kernel (small dot) with dressed internal heavy-meson propagators (double red lines). \textbf{1b.} Dyson equation for the dressed heavy-meson propagator in Eq.~(\ref{eq:mesonprop}). \textbf{1c.} Heavy-meson self-energy in Eq.~(\ref{eq:selfenergy1}). The heavy meson is dressed by the unitarized interaction with pions.}
\end{figure*}

The thermal effects on the unitarized scattering amplitude $\mathcal{T}^{ij}(s)$ from an incoming channel $i$ to an outgoing channel $j$ are then obtained by solving the on-shell Bethe-Salpeter equation in a full coupled-channel basis, with the $s$-wave interaction kernel of Eq.~(\ref{eq:potential}) and the thermal two-meson propagator of Eq.~(\ref{eq:loop}) (see Fig.~\ref{fig:diagrams-a}):
\begin{align}\label{eq:bethesalpeter}
    \mathcal{T}^{ij}(E,\bm{p};T)=  \mathcal{V}^{ij}(s) +\mathcal{V}^{ik}(s)\mathcal{G}^k(E,\bm{p};T)\mathcal{T}^{kj}(E,\bm{p};T) \ .
\end{align}

The spectral functions dressing the meson propagators in Eq.~(\ref{eq:loop}) take into account the modifications due to the presence of interactions with the medium. At finite temperature the heavy meson retarded propagator is defined by
\begin{equation}\label{eq:mesonprop}
 \mathcal{D}_H(\omega,\bm{q}\,;T)=\frac{1}{\omega^2-\bm{q}^2-m_H^2-\Pi_H(\omega,\bm{q}\,;T)} \ ,
\end{equation}
where $m_H$ is the mass of the heavy meson in the vacuum, renormalized by the vacuum contribution of the retarded self-energy $\Pi_H$ (see Fig.~\ref{fig:diagrams-b}). For the purpose of our calculations, using the vacuum propagator for the light meson and thus a $\delta$-type spectral function is a good approximation, as we discussed in our previous works~\cite{Montana:2020lfi,Montana:2020vjg}.

The light-meson contribution to the self-energy of the heavy meson can be obtained by closing the light-meson line in the corresponding $\mathcal{T}$-matrix element (see Fig.~\ref{fig:diagrams-c}), i.e. by integrating over the light-meson four-momenta $q'^\mu=(E',\bm{q}')$. In the ITF it is defined as
\begin{align}\label{eq:selfenergy1}
& \Pi_H(i\omega_n,\bm{q}\,;T)=\nonumber  \\   &-\frac{1}{\beta}\int \frac{d^3q'}{(2\pi)^3} \ \sum_m\mathcal{D}_\Phi(i\omega_m-i\omega_n,\bm{q}\,^\prime) \mathcal{T}_{H\Phi}(i\omega_m,\bm{p}\,) \ .
\end{align}
Using the spectral Lehmann representation for the light meson propagator and the $\mathcal{T}$ matrix, and performing the summation over the Matsubara frequencies $\omega_m$ of the internal $H\Phi$, it reads
\begin{widetext}
\begin{align}\label{eq:selfenergy2}
    \Pi_{H}(i\omega_n,\bm{q}\,;T)&= 
    \frac{1}{\pi} \int\frac{d^3q'}{(2\pi)^3} \int 
    dE \int d\omega
    \frac{\mathcal{S}_\Phi(\omega,\bm{p}-\bm{q}\,)\left[f(E,T)-f(\omega,T)\right]}{E-i\omega_n-\omega} 
    \ \times
    \textrm{Im\,}\mathcal{T}_{H\Phi}(E,\bm{p}\,;T) \ .
\end{align}
\end{widetext}


We note that the self-energy entering in Eq.~(\ref{eq:mesonprop}) can only contain thermal corrections after mass renormalization. However, the self-energy computed with Eq.~(\ref{eq:selfenergy2}) contains both vacuum and thermal corrections. We regularize it by dropping the vacuum contribution, which is identified with the expression obtained when taking the limit $T\to 0$ of Eq.~(\ref{eq:selfenergy2}). See \cite{Montana:2020vjg} for details.

Finally, the spectral function necessary to dress the heavy meson in the two-meson propagator is computed from the imaginary part of the retarded meson propagator,
\begin{equation} \label{eq:specfunc}
  \mathcal{S}_{H}(\omega,\bm{q}\,;T)=-\frac{1}{\pi}\textrm{Im\,}\mathcal{D}_{H}(\omega,\bm{q}\,;T) \ .
\end{equation}

Equations~(\ref{eq:loop}) to (\ref{eq:specfunc}) are interrelated to each other. As a result, solving this set of coupled equations requires an iterative approach until self-consistency is achieved. This process is outlined in Figs.~\ref{fig:diagrams-a}, \ref{fig:diagrams-b}, and \ref{fig:diagrams-c}, where the $\mathcal{T}$-matrix amplitude is depicted as a hatched blob, the perturbative amplitude $\mathcal{V}(s)$ is represented by a small dot, and the propagator of the heavy meson dressed by the medium is shown with double red lines.

\subsection{Non-equilibrium properties}\label{sec:nonequilibrium}
Since heavy mesons have large masses compared to the surrounding light mesons, they are unlikely to achieve thermalization in heavy-ion collisions. Therefore, their time evolution is typically described by a Fokker-Planck (or Langevin) or a Boltzmann approach. The essential components of these approaches are transport coefficients, which can be calculated from the scattering amplitudes of heavy-flavor mesons with light mesons in the hadronic gas. These transport coefficients are typically derived assuming that the light scattering partners are in thermal equilibrium, and they are often calculated as functions of temperature and momentum.

In Ref.~\cite{Torres-Rincon:2021yga} we calculated the transport coefficients of $D$ mesons in the hadronic phase incorporating medium corrections to the scattering amplitudes. To do so, we extended the kinetic theory of $D$ mesons using the more general Kadanoff-Baym equations, so as to account for thermal and off-shell effects. This off-shell kinetic theory also applies to describe the propagation of $\bar{B}$ mesons, as it is valid for any heavy species that can be treated as Brownian particles propagating in a medium of light mesons. In fact, the separation of scales between the heavy-meson mass and the other scales in the system that is exploited to convert the off-shell kinetic equation into a Fokker-Planck equation is larger for the $\bar{B}$ meson than for the $D$ meson. In addition, the quasiparticle approximation that we showed to be sound for the $D$ mesons is even better for $\bar{B}$, since their thermal width is of the same order as that of the $D$ mesons, but their mass is considerably larger, as we will see in Sec.~\ref{sec:results}.

Let us summarize our main results. For a detailed derivation, we recommend to consult our previous work~\cite{Torres-Rincon:2021yga} and references therein. Starting from the Kadanoff-Baym equations and performing a Wigner transform along with a gradient expansion \cite{kadanoff1962quantum}, we arrive to the following form of the off-shell transport equation for the time ordered Green's function of the heavy meson $\mathcal{G}_H^<(X,k)$,
\begin{align}
\Bigg( &  k^\mu - \frac{1}{2} \frac{\partial \textrm{Re } \Pi^{\textrm{R}} (X,k)}{\partial k_\mu} \Bigg) \frac{\partial }{\partial X^\mu}i \mathcal{G}_H^< (X,k) =\nonumber \\ & \frac{1}{2}  i\Pi^< (X,k)  i \mathcal{G}_H^> (X,k) -\frac{1}{2} i\Pi^> (X,k) i \mathcal{G}_H^< (X,k) \ .
\label{eq:transport-eq} 
\end{align}
The lesser and greater Green's functions and the self-energies in Eq.~(\ref{eq:transport-eq}) are functions of the center-of-mass coordinate $X=(t,\bm{X})$ and the four-momentum $k=(k^0,\bm{k})$ of the external heavy meson. Note that $k^0$ and $\bm{k}$ are independent variables, although related through the non-equilibrium spectral function $\mathcal{S}_H(X,k)$. 
Hence the reason we denote this kinetic equation to be ``off-shell'', as the heavy meson is not on its mass shell. 
The self-energy $\Pi^{\textrm{R}}_H(X,k)$ is the extension of the retarded self-energy of Eq.~(\ref{eq:selfenergy1}) to the non-equilibrium case, and the lesser $\Pi^< (X,k)$ and greater $\Pi^> (X,k)$ self-energies can be written in terms of the (retarded) $\mathcal{T}$ matrix of Eq.~(\ref{eq:bethesalpeter}) in the so-called $\mathcal{T}$-matrix approximation~\cite{kadanoff1962quantum,Danielewicz:1982kk,Botermans:1990qi}. 
Inserting appropriate definitions of these quantities, Eq.~(\ref{eq:transport-eq}) can be written in the following form:
\begin{align}\label{eq:transport-eq2}
  \Bigg( &  k^\mu - \frac{1}{2} \frac{\partial \textrm{Re } \Pi^{\textrm{R}}(X,k)}{\partial k_\mu} \Bigg) \frac{\partial}{\partial X^\mu}i \mathcal{G}_H^< (X,k)  =\nonumber \\ & \frac12\int \frac{dk_1^0}{2\pi} \frac{d^3q}{(2\pi)^3} \left[  W(k^0,\bm{k}+\bm{q}, k_1^0,\bm{q}\,) i \mathcal{G}_H^< (X,k^0,\bm{k}+\bm{q}\,)\right. \nonumber \\ & \quad \left.  -\, W(k^0,\bm{k},k_1^0,\bm{q}\,) i \mathcal{G}_H^<(X,k^0,\bm{k}\,)\right] \ .   
\end{align}
The off-shell collision rate of a heavy meson with energy $k^0$ and momentum $\bm{k}$ to a final heavy meson with energy $k_1^0$ and momentum $\bm{k}-\bm{q}$, with momentum loss $\bm{q}\equiv\bm{k}-\bm{k}_1$, is defined as
\begin{widetext}
\begin{align}
W (k^0,\bm{k},k_1^0,\bm{q}\,) & \equiv \int \frac{d^4k_3}{(2\pi)^4} \frac{d^4k_2}{(2\pi)^4} \ (2\pi)^4 \delta (k_1^0+k_2^0-k_3^0-k^0) \delta^{(3)} (\bm{k}_2-\bm{k}_3-\bm{q}\,) 
\  \left|\mathcal{T} (k_1^0+k_2^0+i\varepsilon, \bm{k} - \bm{q} + \bm{k}_2)\right|^2  \nonumber \\
& \times \ i \mathcal{G}_\Phi^>(X,k_2) \ i \mathcal{G}_\Phi^<(X,k_3) \ i \mathcal{G}_H^>(X,k_1^0,\bm{k}-\bm{q}\,) \ . \label{eq:transport-scatrate}
\end{align}
\end{widetext}
The labels of the momenta correspond to the choice of a generic scattering process $H(k)+\Phi(k_3)\to H(k_1)+\Phi(k_2)$.
Each of the two terms of the right-hand side of Eq.~(\ref{eq:transport-eq2}) can be identified as the collision gain and loss terms respectively. It is important to note that in Eq.~(\ref{eq:transport-scatrate}) there is an implicit sum over the different species $\Phi$ and $H$ that can interact with the external off-shell heavy meson. 

Next, one may exploit the separation of scales between the meson masses to arrive to an off-shell Fokker-Planck equation for $i\mathcal{G}_H^<(t,k)$.
While the derivation of a transport equation for heavy mesons required the use of real-time techniques, the actual calculation of the heavy-flavor transport coefficients can be addressed in a near-equilibrium regime, in which the temperature is at least locally well defined. While thermal local equilibrium can be safely considered for the light mesons, it is also reasonable to assume that the heavy mesons are not far from equilibrium. Then Eq.~(\ref{eq:transport-eq2}) can be written as follows:
\begin{widetext}
\begin{align} 
\frac{\partial}{\partial t} i \mathcal{G}_H^< (t,k) = \frac{\partial}{\partial k^i} \Bigg\{ \hat{A} (k;T) k^i\, i \mathcal{G}_H^< (t,k) + \frac{\partial}{\partial k^j} \left[ \hat{B}_0(k;T) \Delta^{ij} + \hat{B}_1(k;T) \frac{k^i k^j}{\bm{k}\,^2} \right] i \mathcal{G}_H^< (t,k) \Bigg\} \, , \label{eq:transport-offFP} 
\end{align}
\end{widetext}
with $\Delta^{ij}=\delta^{ij}-k^ik^j/\bm{k}^2$, and the  transport coefficients $\hat{A}(k^0,\bm{k};T)$, $\hat{B}_0(k^0,\bm{k};T)$, and $\hat{B}_1(k^0,\bm{k};T)$ defined off shell and at temperature $T$. The drag force coefficient is given by
\begin{equation}
     \hat{A} (k^0, \bm{k}\,;T) \equiv \left \langle 1 -\frac{\bm{k} \cdot \bm{k}_1}{\bm{k}\,^2} \right \rangle \ , \label{eq:transport-hatA} \\
\end{equation}
and the transverse and longitudinal momentum diffusion coefficients read
\begin{subequations}
\begin{align}
 \hat{B}_0 (k^0, \bm{k}\,;T) & \equiv  \frac14 \left \langle \bm{k}_1\,^2 - \frac{(\bm{k} \cdot \bm{k}_1)^2}{\bm{k}\,^2} \right \rangle \ , \label{eq:transport-hatB0} \\
  \hat{B}_1 (k^0, \bm{k}\,;T) & \equiv \frac12 \left \langle  \frac{[ \bm{k} \cdot (\bm{k}-\bm{k}_1)]^2}{\bm{k}\,^2} \right \rangle \ . \label{eq:transport-hatB1}
\end{align}
\end{subequations}
The angle brackets $\langle\mathcal{F}(\bm{k},\bm{k}_1)\rangle$ denote the average of the generic quantity $\mathcal{F}(\bm{k},\bm{k}_1)$, which is defined as 
\begin{widetext}
\begin{align}
 \big\langle \mathcal{F}(\bm{k},\bm{k}_1) \big\rangle & = \frac{1}{2k^0} \sum_{\lambda,\lambda'=\pm} \lambda \lambda' \int_{-\infty}^\infty \ dk_1^0  \int \prod_{i=1}^3 \frac{d^3k_i}{(2\pi)^3} \ \frac{1}{2E_22E_3}  \ S_H(k_1^0,\bm{k}_1;T)    (2\pi)^4 \delta^{(3)} (\bm{k}+\bm{k}_3-\bm{k}_1-\bm{k}_2) \nonumber \\
& \times \delta (k^0+\lambda' E_3- \lambda E_2-k^0_1) \left|\mathcal{T}(k^0+ \lambda' E_3,\bm{k}+\bm{k}_3;T)\right|^2   \ f_\Phi (\lambda'E_3;T) \tilde{f}_\Phi (\lambda E_2;T)  \tilde{f}_H (k_1^0;T) \ \ \mathcal{F}(\bm{k},\bm{k}_1)  \ , \label{eq:transport-rateoff}
\end{align}
\end{widetext}
where $f(E_i;T)$ is the equilibrium occupation number, i.e. the Bose-Einstein distribution function, and $\tilde{f}(E_i;T)=1+f(E_i;T)=-f(-E_i;T)$ is the Bose enhancement factor.
Equation~(\ref{eq:transport-rateoff}) incorporates the equilibrium quantities presented in the previous section, i.e. the equilibrium thermal scattering amplitudes and spectral functions. In particular, it is a sum of four terms ($\lambda,\lambda'=\pm$) that evaluate the $\mathcal{T}$ matrix in different energy regions. The relevance of the contribution of each of these terms to the transport coefficients is discussed in the next section.

\section{Results}\label{sec:results}

The formalism described in the previous section provides a framework to compute the in-medium properties of heavy mesons. Here we present our results for $D$ and $\bar{B}$ mesons for temperatures below the deconfinement transition temperature $T_c\sim 155$ MeV in HICs~\cite{Borsanyi:2010bp,Bazavov:2011nk}.
This applicability limit of our approach is inherent to the effective theory with hadronic degrees of freedom and massive Goldstone bosons upon which it is built. Therefore it is important to be careful when interpreting the results at our highest temperatures $T\sim 150$~MeV, as the system will begin to transition into the deconfined phase. Additionally, while the unitarized version of HMChPT extends the validity of the low-energy theory to higher energies, for temperatures exceeding $T\sim 150$~MeV, the thermal energies of the mesons may fall outside the energy region of applicability of the theory, as we noted in Ref.~\cite{Montana:2020lfi}.

\begin{figure*}[htbp!]
\begin{center}
\includegraphics[width=\textwidth]{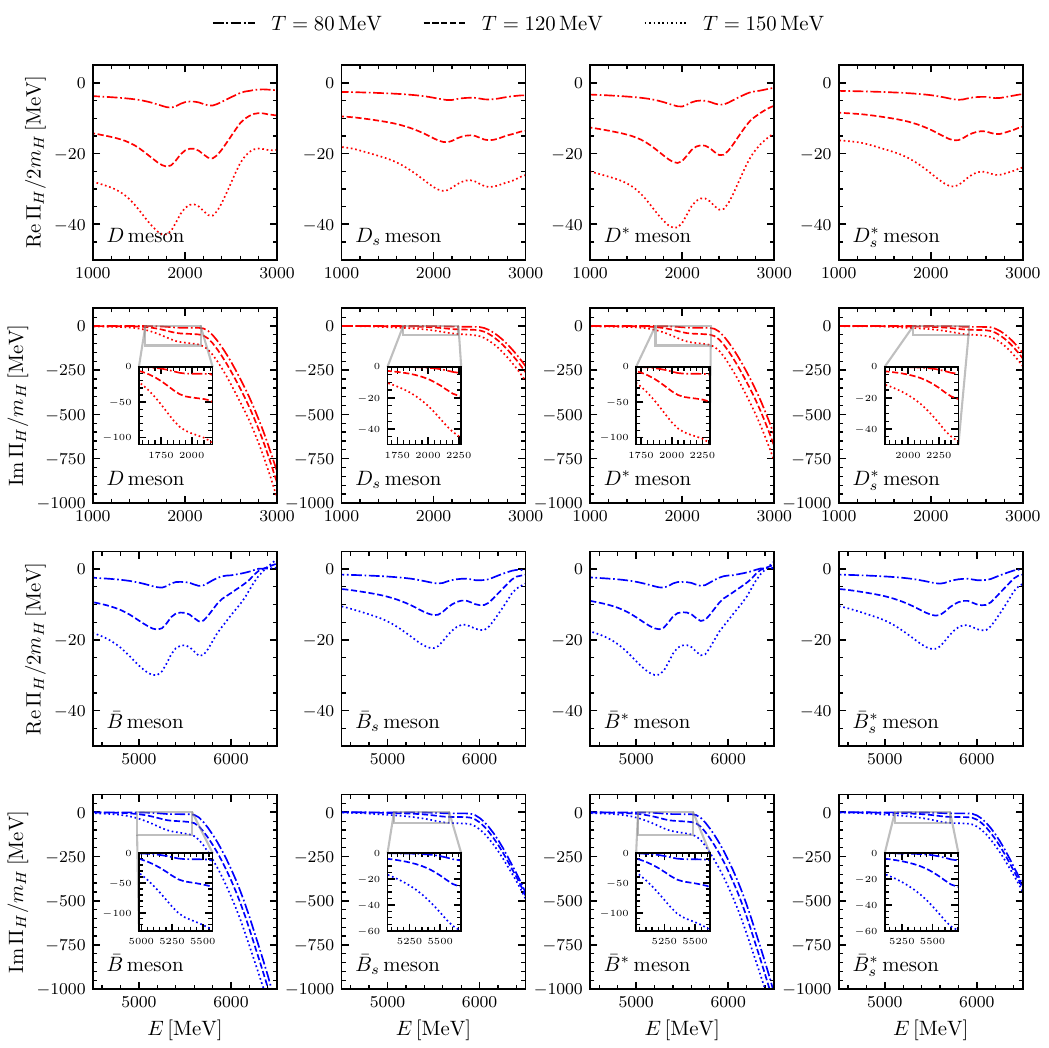}
\end{center}
\caption{Real and imaginary parts of the pion contribution to the self-energy of the ground-state heavy mesons at several temperatures (see legend). Panels in the two top rows correspond to charmed mesons, and panels in the two bottom rows to bottomed mesons. Different columns correspond to states with angular momentum and strangeness $(J,S)=(0,0),\,(0,1),\,(1,0),\,(1,1)$, in this order.
\label{fig:selfenergy}}
\end{figure*}

For temperatures $T\lesssim 150$~MeV, pions give the largest contribution to the medium corrections, as being the lightest mesons they are the most abundant species in the thermal bath. Unless otherwise stated, in the calculations presented in this work we only consider the thermal effects due to pions and neglect the contribution of the heavier kaons and eta mesons. We note that the results in the charm sector were already published in our previous works~\cite{Montana:2020lfi,Montana:2020vjg,Torres-Rincon:2021yga} and are reproduced here for the sake of comparison between the bottom and charm sectors.

\subsection{Self-energies}\label{sec:selfenergy}
We start with the discussion of the self-consistent results of the self-energy of the ground-state heavy mesons in a pionic medium at finite temperature, displayed as a function of the energy in Fig.~\ref{fig:selfenergy}, for zero three-momentum and scaled by the mass of the heavy meson in vacuum. We show the results for three different values of the temperature of the medium, $T=80,\, 120$ and $150$~MeV, in different line styles.

The real part of the self-energy is related to the thermal correction to the mass. 
This is evident from the expression of the heavy-meson retarded propagator at finite temperature in Eq.~(\ref{eq:mesonprop}).
In the quasiparticle approximation, which we will see is well-grounded for both $D$ and $\bar{B}$ mesons, and if the thermal propagator's pole is close to the vacuum pole, the mass shift is roughly given by $\Delta m_H\approx {\textrm{Re}}\,\Pi_H(m_H,\bm{0};T)/(2m_H)$. 
Therefore, as shown in the first row panels of Fig.~\ref{fig:selfenergy} for charmed mesons and in the third row for bottomed mesons, the negative character of the real part of the self-energy indicates that the masses of the heavy mesons will decrease as temperature rises. The fact that the real part of the self-energy is more negative for the nonstrange mesons than for the strange mesons is explained by the large attractive interaction in the $D^{(*)}\pi$ and $\bar{B}^{(*)}\pi$ channels after unitarization.
Furthermore, one can see that the values of the real part of self-energy over the heavy meson mass at a particular temperature are similar for $D_{(s)}$ and $D^*_{(s)}$, as well as for $\bar{B}_{(s)}$ and $\bar{B}^*_{(s)}$. Although it appears to be less negative for bottom than for charm, it is of comparable size in both sectors. These findings are closely connected to the HQSFS intrinsic of the interaction. It is also important to note that the quantitative comparison of the results in the two flavor sectors may be impacted by the details of the numerical calculations, e.g. by the choice of the infinite integration limits, or the limitations of the effective theory at high energies.
The authors of Ref.~\cite{Cleven:2017fun} neglected the shift of the in-medium mass of the heavy mesons by setting to zero the real part of the respective self-energies. Although small compared to the vacuum mass, $|\Delta m_H|/m_H\sim 1-2\%$, we consider that it is important and keep the full self-energy for the calculation of the spectral function.

\begin{figure*}[htbp!]
\begin{center}
\includegraphics[width=\textwidth]{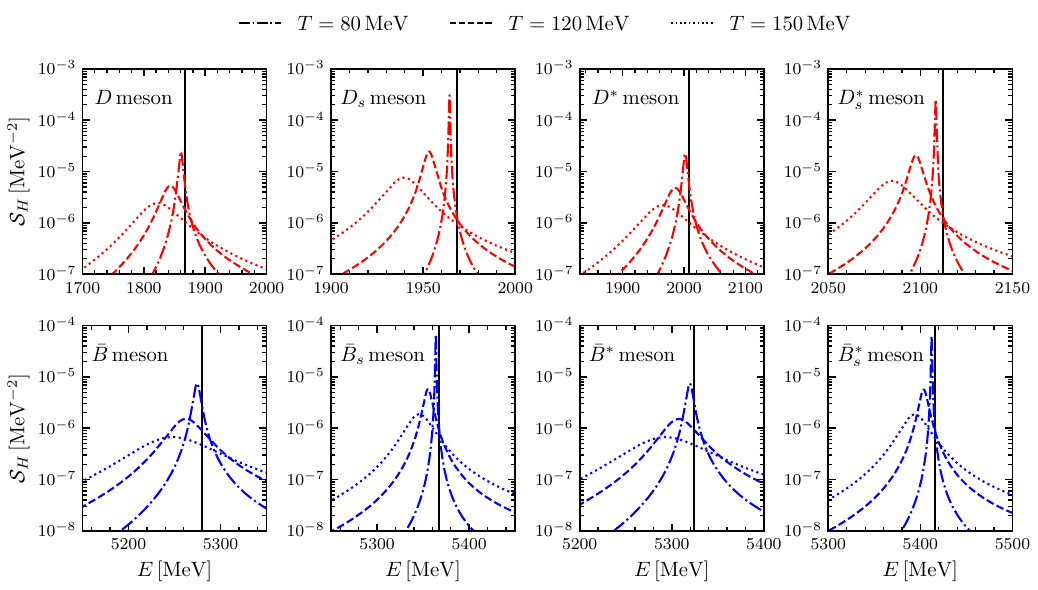}
\end{center}
\caption{Spectral function of the ground-state charmed mesons (top panels) and bottomed mesons (bottom panels) at several temperatures (see legend). The column description is the same as in Fig.~\ref{fig:selfenergy}. Vertical lines depict the values in vacuum ($T=0$).
\label{fig:spectralfunction}}
\end{figure*}

The imaginary part of the self-energy relates to the thermal width acquired by the heavy meson due to interactions with pions within the medium. 
The panels in the second and fourth rows of Fig.~\ref{fig:selfenergy} show the imaginary part of the self-energy of the charmed and bottomed mesons, respectively, over their respective mass in vacuum. The insets provide a zoom in the region $E\approx [m_H-2m_\pi,\,m_H+2m_\pi]$.
Similar features are observed in all the panels. The imaginary part of the self-energy is essentially zero for energies below $m_H-2m_\pi$, above which it starts decreasing mildly.
This initial drop at around $m_H$ is exclusively caused by the presence of the thermal medium, which allows for the absorption of two thermal pions. 
These absorption processes make it possible for the scattering amplitude to be non-zero even below the two-meson threshold due to the so-called Landau cut \cite{Weldon:1983jn,Ghosh:2011bw} of the two-meson propagator (see also our discussion in Refs.~\cite{Montana:2020vjg,Torres-Rincon:2021yga,MontanaFaiget:2022cog}). Our calculations show that also the self-energy of the heavy meson can reveal the effects of the Landau cut, thanks to the self-consistency of our approach.
This effect becomes more relevant at higher temperatures, i.e. $T\gtrsim 100$~MeV, where the pion density is larger.
A substantially larger drop takes place at $m_H + 2m_\pi$, which is the energy where the heavy meson at rest can emit two pions. This later growth of the magnitude of the imaginary part of the self-energy takes place at similar rates for all temperatures, since the emission of two pions is also possible in vacuum for a large enough energy of an off-shell heavy meson, and it is related to the unitary cut of the propagator. As a result of the combination of the Landau and unitary cut effects, and by virtue of the relation between ${\textrm{Im}}\,\Pi_H$ and the thermal decay width, we expect the ground-state spectral functions to broaden as the temperature of the thermal medium increases. Similarly as it happened for the real parts, the magnitude of ${\textrm{Im}}\,\Pi_H/m_H$ at a given temperature is similar in size when comparing results for pseudoscalar and vector mesons, and it is somewhat larger in the bottom sector than in the charm sector. 

\subsection{Spectral functions}\label{sec:spectralfunction}
The spectral function for the ground-state heavy mesons follows the standard definition in terms of the retarded propagator---see Eqs.~(\ref{eq:mesonprop}) and (\ref{eq:specfunc})---, in which the self-energy is responsible for the thermal corrections with respect to the vacuum propagator. In Fig.~\ref{fig:spectralfunction} we show the energy dependence of the spectral function of the charmed mesons (top panels) and the bottomed mesons (bottom panels) at rest, at the same temperatures as for the self-energy described above. The vertical solid lines depict the corresponding value of the mass in vacuum. From these plots, the drop of the mass and the increase of the width anticipated from analyzing the self-energies become manifest. This is evident as the maximum of the spectral function shifts towards lower energies and it becomes wider with increasing temperature.

\begin{figure*}[hbtp!]
\begin{center}
\includegraphics[width=0.7\textwidth]{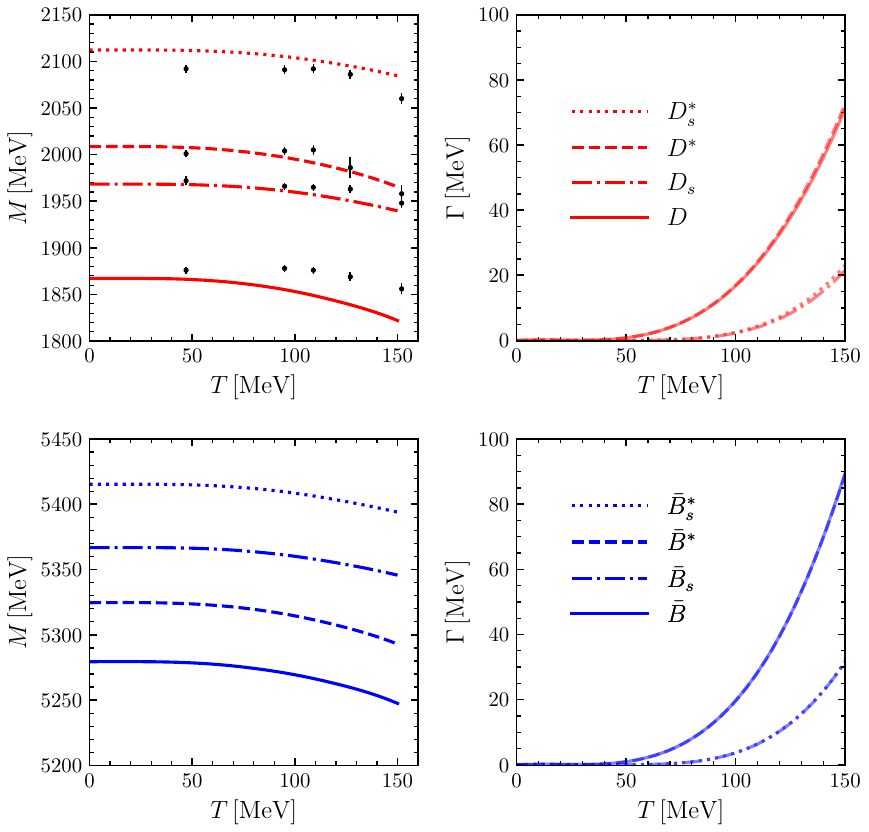}
\end{center}
\caption{Temperature evolution of the mass (left panels) and the width (right panels) of ground-state charmed mesons (top) and bottomed mesons (bottom). Data points in the top left panel correspond to lattice QCD calculations from Ref.~\cite{Aarts:2022krz}.
\label{fig:widths}}
\end{figure*}

In the quasiparticle approximation, the spectral function admits a Lorentzian shape peaked at the quasiparticle energy $E_k(T)$ (with $M(T)\equiv E_k$ at rest) and a spectral width $\gamma_k(T)\ll E_k(T)$.
For the spectral functions in Fig.~\ref{fig:spectralfunction}, which are narrow, the quasiparticle approximation is indeed justified.
Figure~\ref{fig:widths} shows the values of the mass (left panels) and the decay width (right panels) as a function of the temperature determined by analyzing the position and the width of the peak of the spectral functions. For the charmed mesons, we find a reduction in mass $\approx 45$~MeV and $\approx 25$~MeV for the non-strange and the strange states, respectively, at the highest temperature $T=150$~MeV, and corresponding thermal widths of $\approx 70$~MeV and $\approx 20$~MeV. For the respective bottomed mesons, the reduction in mass is $\approx 30$~MeV and $\approx 20$~MeV, and the acquired width is $\approx 90$~MeV and $\approx 30$~MeV. The thermal masses of the $D^{(*)}$ and $D_s^{(*)}$ mesons were calculated using lattice QCD simulations in Ref.~\cite{Aarts:2022krz}. We show their results in the top left panel of Fig.~\ref{fig:widths}, although for the comparison with our calculations one has to keep in mind that a systematic shift is to be expected due to the use of heavier than physical pions in the lattice. Indeed, smaller thermal modifications at a given temperature are consistent with the lower abundances of heavier pions.

\begin{figure*}[htbp!]
\begin{center}
\includegraphics[width=0.37\textwidth]{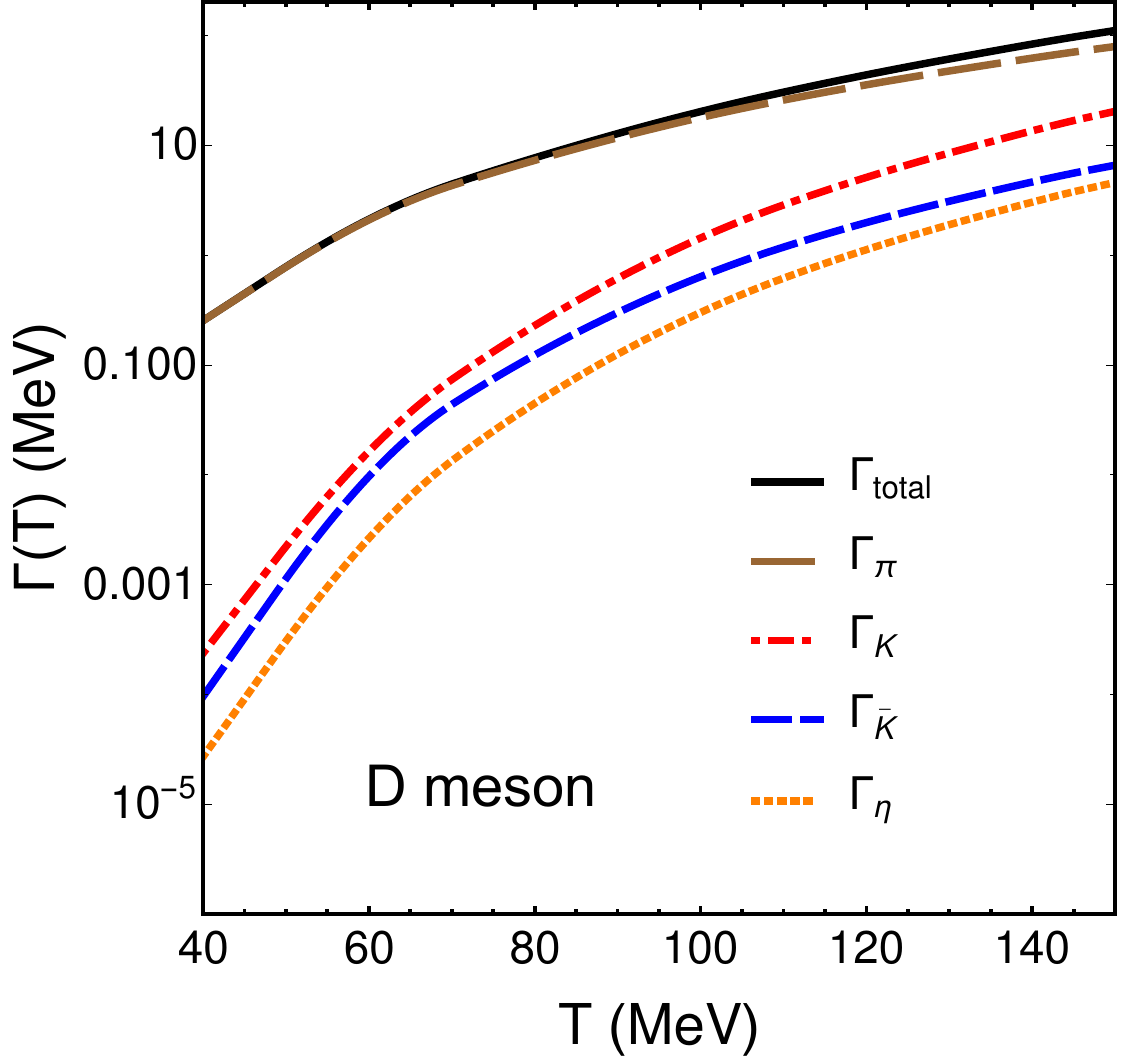}\hspace{0.5cm}
\includegraphics[width=0.37\textwidth]{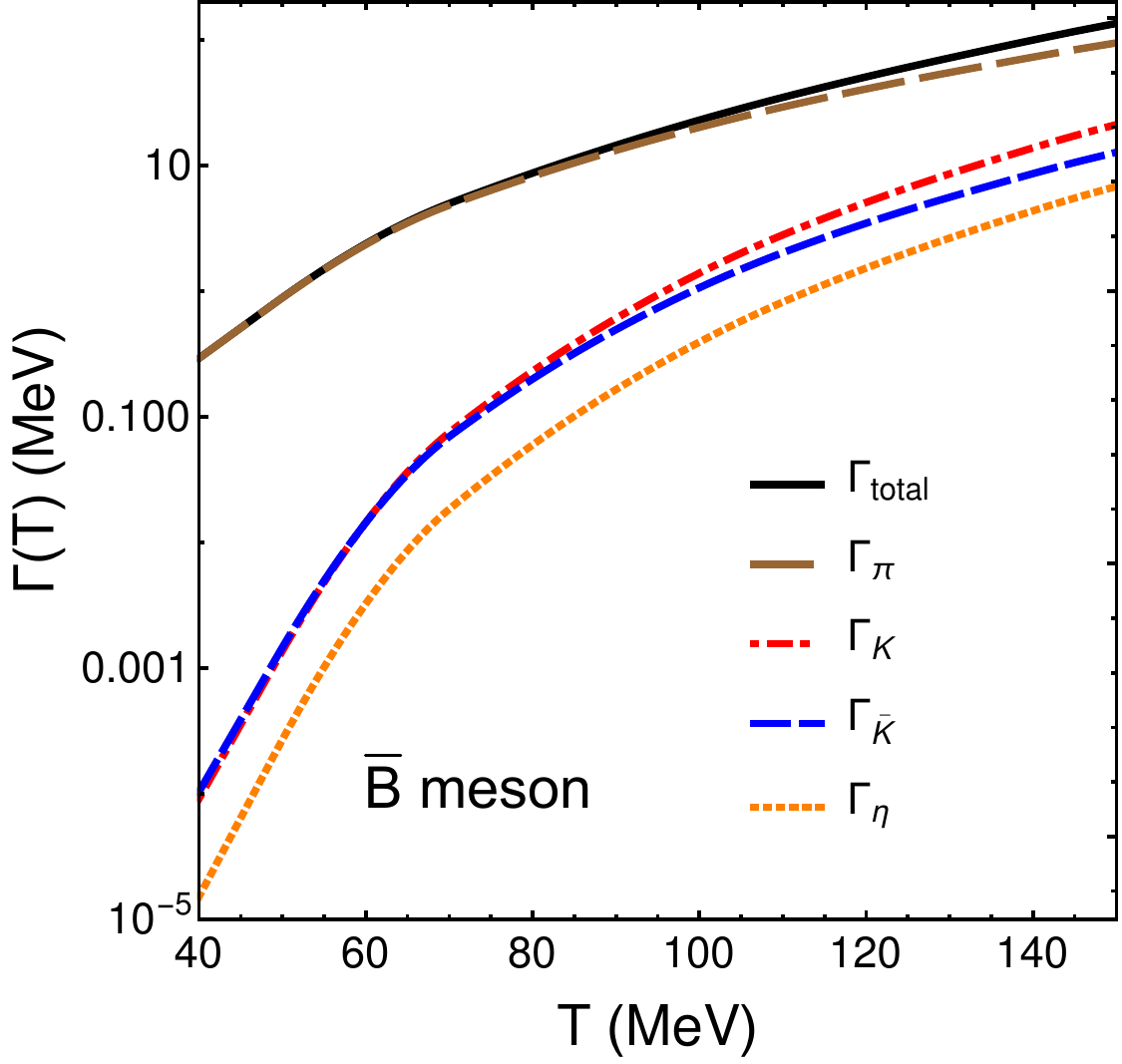}
\end{center}
\caption{Contribution to the averaged thermal width of the $D$ meson (left panel) and the $\bar{B}$ meson (right panel) from a thermal bath of pions, kaons, antikaons, and $\eta$ mesons.
\label{fig:transport-widths}}
\end{figure*}

\begin{figure*}[htbp!]
\begin{center}
\includegraphics[width=\textwidth]{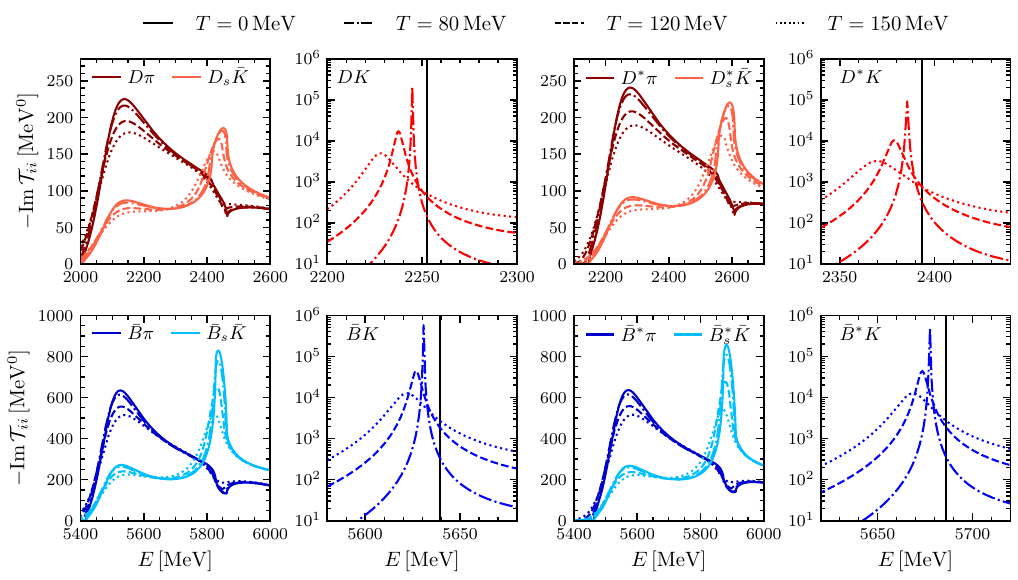}
\end{center}
\caption{Imaginary part of the diagonal elements of the scattering amplitudes in the charm sector (top panels) and in the bottom sector (bottom panels) at several temperatures. Different columns correspond to states with angular momentum, strangeness, and isospin $(J,S,I)=(0,0,1/2),\,(0,1,0),\,(1,0,1/2),\,(1,1,0)$, in this order. Vertical lines in the sectors with strangeness depict the energy location of the bound states in vacuum.
\label{fig:ImT}}
\end{figure*}

We remind that in the self-consistent calculations of the self-energies and the spectral functions, only the impact of the thermal pions is taken into account, arguing the contribution of other light mesons to be presumably suppressed. The contribution of each of the light mesons in the medium to the thermal width of the heavy meson can be easy analyzed from its definition in terms of the retarded self-energy in the quasiparticle approximation,
\begin{equation}
    \Gamma_k=-\frac{z_k}{E_k}{\rm Im\,}\Pi(E_k,\bm{k};T),
\end{equation}
with $z_k\approx 1$. Since ${\rm Im\,}\Pi(E_k,\bm{k};T)$ is given by the integration over the imaginary part of the unitarized scattering amplitude, ${\rm Im\,}\mathcal{T}$, and we have access to all the matrix elements $\mathcal{T}_{ij}$, we can readily assess the effect of the four elastic channels for the interactions of a heavy meson with the light pseudoscalars ($\pi,\,K,\,\bar{K},\,\eta$). We showed in Ref.~\cite{Torres-Rincon:2021yga} that the effect of the inelastic channels is negligible. In Fig.~\ref{fig:transport-widths} we show, in logarithmic scale, the contribution to the width of the $D$ meson (left) and the $\bar{B}$ meson (right) coming from the different light mesons, averaged over momenta~\cite{Torres-Rincon:2021yga}. At low temperatures, the kaons and the $\eta$ mesons have a negligible contribution because of their small abundances, as expected. The only relevant contribution is that of the pions. Close to $T = 150$~MeV, the more massive mesons contribute several MeV to the heavy-meson decay width, but are still subdominant compared to the pion.

The process of unitarization of the scattering amplitude of Eq.~(\ref{eq:bethesalpeter}) leads to the emergence of two poles in the sectors with strangeness $S=0$ and isospin $I=1/2$ that correspond to the two-pole structure of the $D_0^*(2300)$, in the case of total angular momentum $J=0$, and of the $D_1(2430)$, in the case of $J=1$. The same applies for the counterparts in the bottom sector. In the sectors with $(S,I)=(1,0)$, the poles of the $D_{s0}^*(2317)$ and the $D_{s1}(2460)$ emerge for $J=0$ and $J=1$, respectively, as well of their bottomed analogues.
The characterization of these states requires the analytical continuation of $\mathcal{T}$ to complex energies. The pole position in the complex-energy plane provides the mass and the half-width of these states. While this is a well-established procedure at $T=0$, the poles search at finite temperature is a complex task for two reasons. Firstly, one has to deal with the analytic continuation of imaginary frequencies, and secondly, a numerical search of a singularity in the complex plane within self-consistency is computationally challenging.
Alternatively, to determine the spectral properties of the dynamically generated states at finite temperature, we use the imaginary part of $\mathcal{T}$ on the real axis, shown in Fig.~\ref{fig:ImT}, as a proxy for their spectral shape. From the several coupled channels in each sector, we choose to plot the diagonal element $\mathcal{T}_{ii}$ for the channel $i$ to which the state couples more strongly in vacuum. The numerical values of the vacuum properties and effective couplings are given in our previous works~\cite{Montana:2020vjg,MontanaFaiget:2022cog}. 

In the cases with zero strangeness, the proximity of the position of the resonances to channel thresholds gives rise to complicated structures. However, one can still clearly recognize a gradual evolution of the peaks and widths as $T$ increases. The strange sectors are more straightforward. The $T=0$ delta-type spectral function (i.e. bound state) acquires a non-zero width, and the shift and widening of the peak is comparable to that of the ground states. Nevertheless, an increase in strength is visible on the right-hand side of the distributions. This asymmetry can be explained by the fact that the channel threshold is not sharp anymore due to the widening of the $D^{(*)}$ or $\bar{B}^{(*)}$ meson, and it is lowered in energy due to the decrease of the heavy-meson mass with temperature. 
Both of these effects open the phase space for decay into this channel at lower energies.

\begin{figure*}[htbp!]
\begin{center}
  \includegraphics[width=0.35\textwidth]{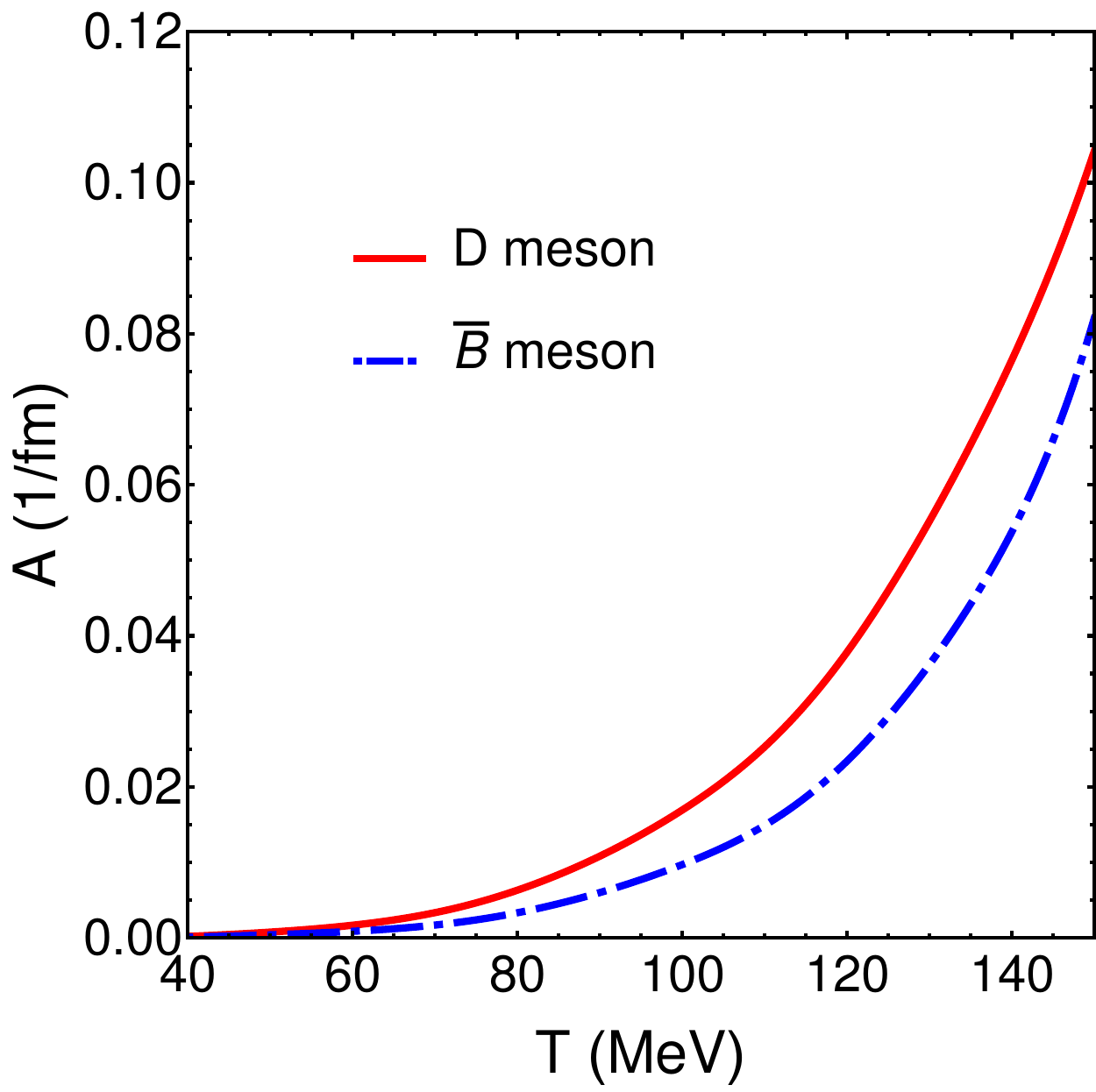}\hspace{0.5cm}
  \includegraphics[width=0.35\textwidth]{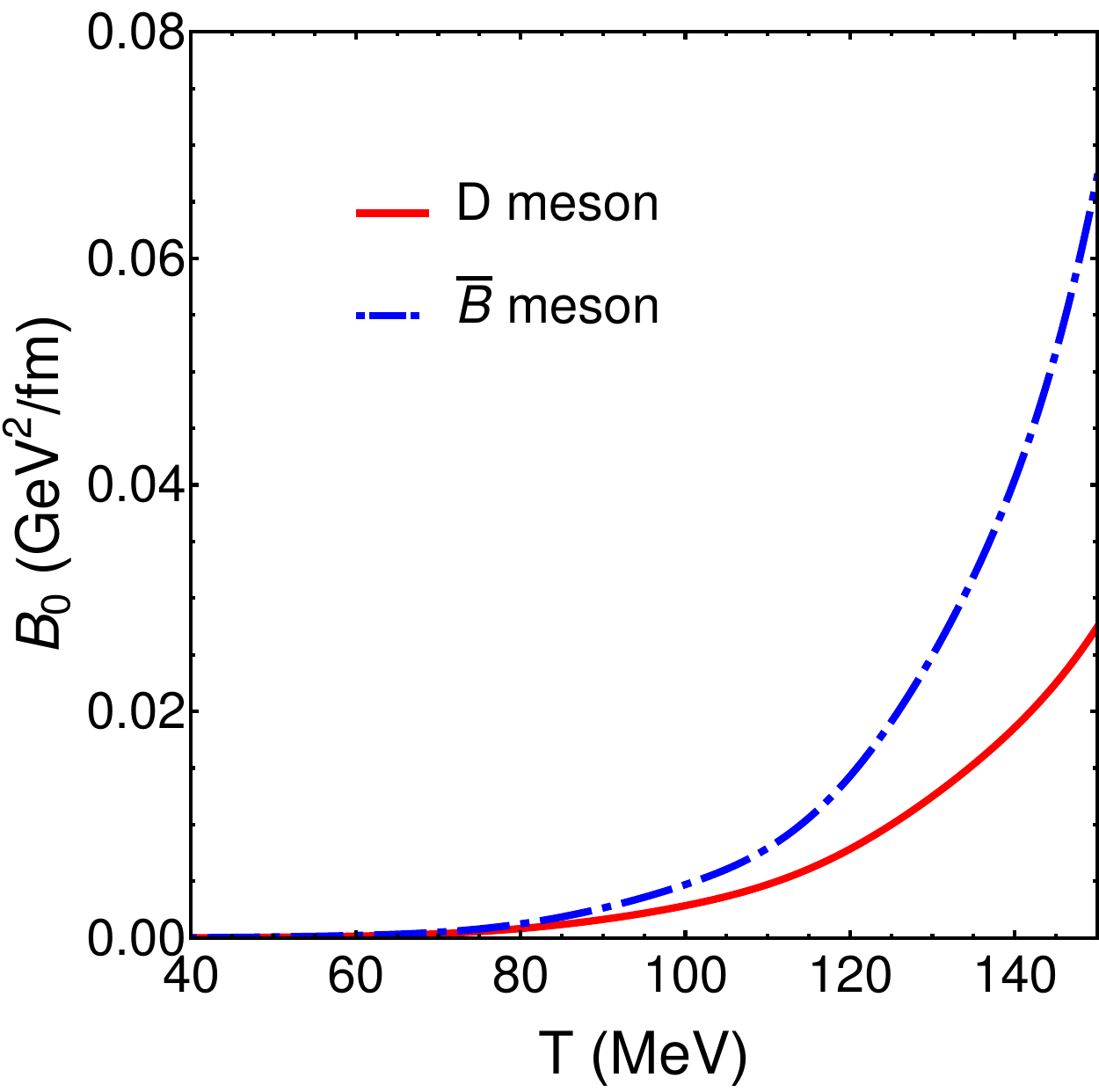}
\end{center}
\caption{Transport coefficients of the $\bar{B}$ meson in the static limit $\bm{k}\rightarrow \bm{0}$ (where $B_1=B_0$), compared to the results for the $D$ meson. }
\label{fig:transport-A-B}
\end{figure*}

\subsection{Transport coefficients}\label{sec:transportcoef}

Now, we discuss the results of the transport coefficients for a heavy meson propagating through the hadronic medium. We will specifically focus on the comparison between $D$ and $\bar{B}$ mesons. We start with two of the transport coefficients defined in momentum space, the drag force in Eq.~(\ref{eq:transport-hatA}) and the transverse diffusion coefficient of Eq.~(\ref{eq:transport-hatB0}). To reduce the number of variables we will present the results in the so-called static limit $\bm{k} \to \bm{0}$, i.e. for a low-momentum heavy meson. In the off-shell version of the Fokker-Planck equation~(\ref{eq:transport-offFP}), these transport coefficients also depend on the value of $k^0$. Since the quasiparticle approximation is excellent for the temperatures considered, we will set $k^0$ to the quasiparticle mass, that is $k^0=E_{k \rightarrow 0}=M(T)$. Then, the only remaining dependence is on the temperature. In the left panel of Fig.~\ref{fig:transport-A-B} we present the comparison of the drag force coefficient for the two heavy flavors as functions of temperature, while the right panel displays the transverse diffusion coefficients. In the static limit, we have checked that the longitudinal diffusion coefficient $B_1(k^0, \bm{k} \rightarrow \bm{0};T)$ is degenerate with $B_0(k^0, \bm{k} \rightarrow \bm{0};T)$. As explained in Ref.~\cite{Torres-Rincon:2021yga}, these coefficients can be computed with different degrees of approximation, but in this work we only present the complete off-shell computation. This calculation incorporates: 1) the exact thermal spectral function of the heavy meson as required in the average of Eq.~(\ref{eq:transport-rateoff}); 2) the full thermal $\mathcal{T}$-matrix appearing in the same equation; 3) quantum effects encoded in the Bose enhancement factors; and 4) all kinematic processes allowed by energy-momentum conservation, including number-conserving  
($2 \leftrightarrow 2$) and number-violating processes ($1 \leftrightarrow 3$). 

The $2 \leftrightarrow 2$ scatterings are described by the $\lambda=\lambda'$ terms in Eq.~(\ref{eq:transport-rateoff}), while the $1 \leftrightarrow 3$ processes take $\lambda=-\lambda'$ This fact can easily be grasped by looking at the signs of the energy conservation delta. In Ref.~\cite{Torres-Rincon:2021yga} we reported that the number-violating processes contribute very little to the transport coefficients, while the $2 \leftrightarrow 2$ collisions make the leading contributions. Among the latter, the case $\lambda=\lambda'=+$ corresponds to the standard term in which the binary collision is taking place at an energy corresponding to the sum of the incoming energies $k^0+E_3$ [cf. Eq.~(\ref{eq:transport-rateoff})]. However the case $\lambda=\lambda'=-$, corresponds to a binary collision in which the scattering matrix is evaluated at the energy difference $k^0-E_3$. For typical energies around the quasiparticle masses, this difference probes the kinematic region below the two-particle threshold. For interactions computed in vacuum, this region has a vanishing $\mathcal{T}$-matrix amplitude, and this entire process can be safely neglected. However, for interactions self-consistently calculated at $T\neq 0$ the $\mathcal{T}$-matrix has a nonvanishing support in this region, due to the Landau cuts (see Sec.~\ref{sec:selfenergy}).  We have proven in Ref.~\cite{Torres-Rincon:2021yga} that the contribution of these processes cannot be neglected and it becomes comparable to the contribution stemming from the unitary cut. Eventually, we have obtained transport coefficients that are up to a factor of three larger with respect to previous results at the highest temperatures. We nonetheless agree at low temperatures, where the Landau cut disappears.

To compare the results between $D$ and $\bar{B}$ mesons in Fig.~\ref{fig:transport-A-B} we recall that a simple nonrelativistic approximation for the momentum-space diffusion coefficient $B_0$ has no leading-order dependence on the heavy mass~\cite{Abreu:2012et}, but a dependence proportional to the total cross section, $B_0 \propto \sigma$.  From Fig.~\ref{fig:ImT} we have learnt that the cross sections (proportional to $\textrm{Im } \mathcal{T}$) of $\bar{B}$ mesons are a factor 2-3 larger than the those of $D$ mesons. This explains why the $B_0$ coefficient is 2-3 times larger for bottom than for charm. On the other hand, the drag force coefficient does have a leading dependence on the heavy mass $M_H$. Again from Ref.~\cite{Abreu:2012et} the nonrelativistic expression goes like $A \propto \sigma/M_H$. Therefore, going from charm to bottom, the gain factor from the cross section is approximately compensated by the reducing factor due to the increasing mass, as $M_D/M_{\bar{B}} \simeq 1/3$. Therefore we expect that $A_{\bar{B} \textrm{ meson}} \lesssim A_{D \textrm{ meson}}$, which is what we observe in the left panel of Fig.~\ref{fig:transport-A-B}.

\begin{figure*}[htbp!]
\begin{center}
  \includegraphics[width=0.35\textwidth]{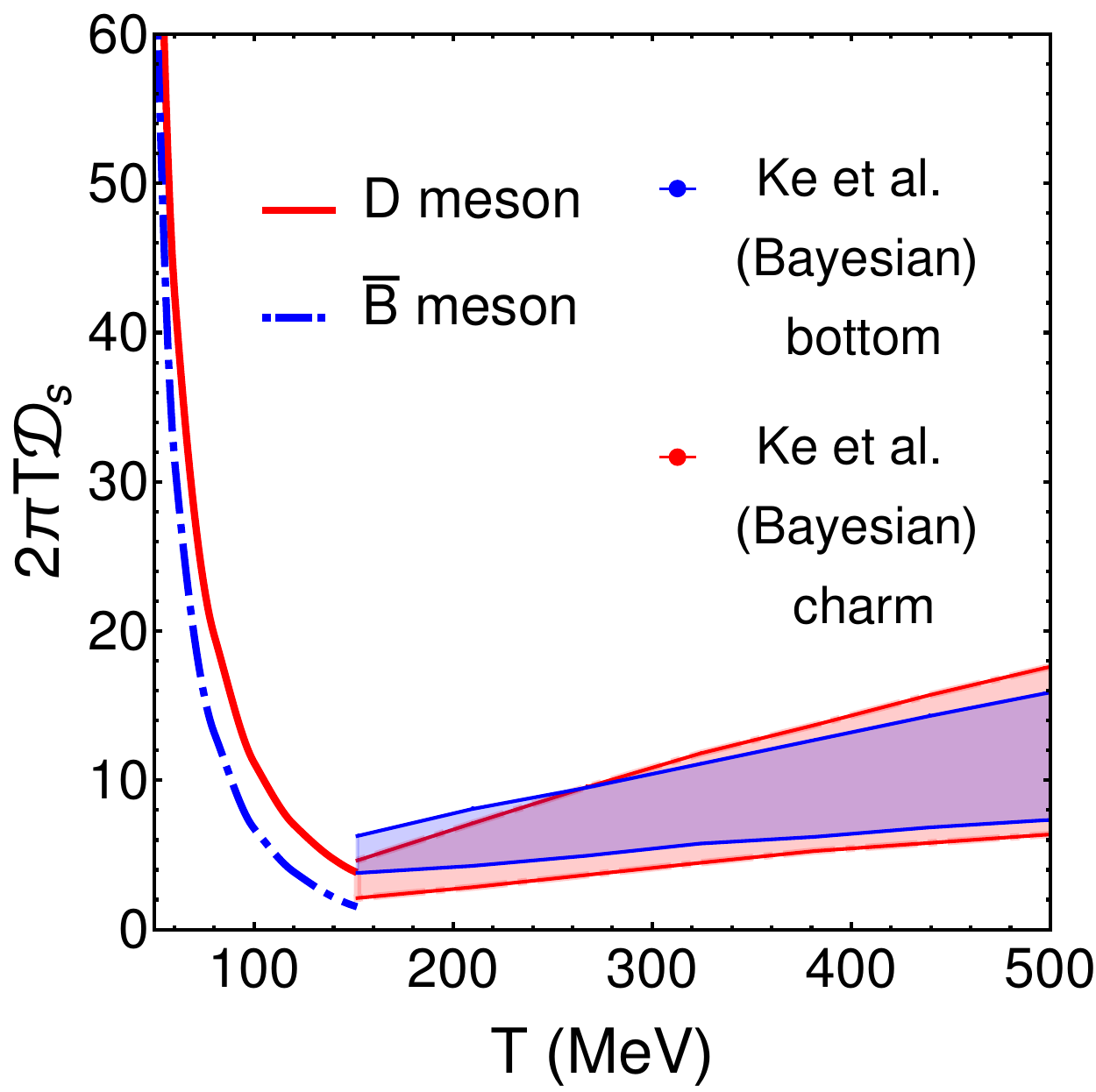}\hspace{0.5cm}
  \includegraphics[width=0.35\textwidth]{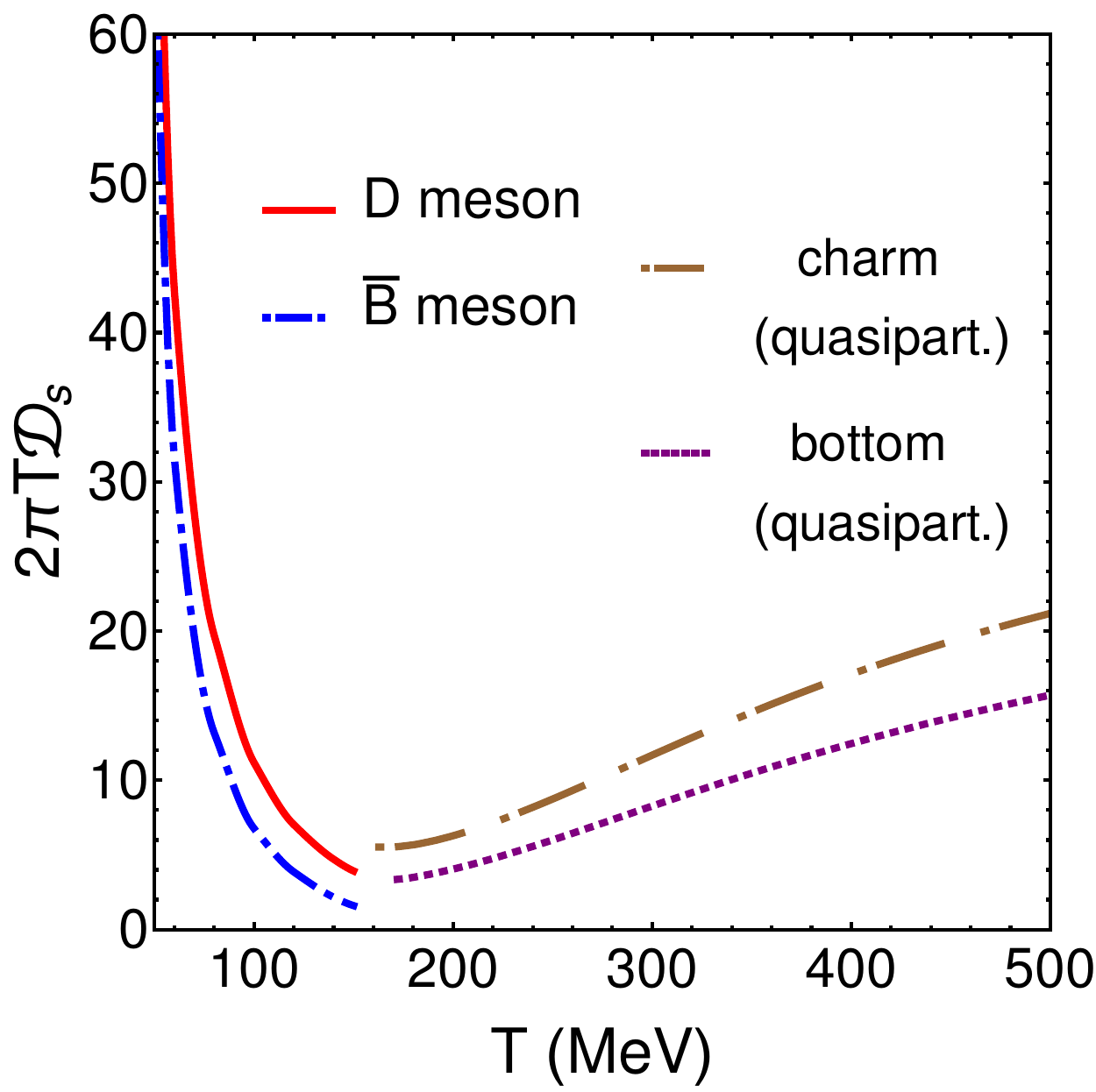}
\end{center}
\caption{Off-shell spatial diffusion coefficient of the $\bar{B}$ meson (normalized by the thermal wavelength) around $T_c$, together with the results for the $D$ meson, and compared to the calculations above $T_c$ from the Bayesian calculation of Ref.~\cite{Ke:2018tsh} (left panel), and from the quasiparticle model of Ref.~\cite{Das:2016llg} (right panel).}
\label{fig:transport-Ds}
\end{figure*}

Finally we plot in Fig.~\ref{fig:transport-Ds} the so-called spatial diffusion coefficient, 
\begin{align}
  \mathcal{D}_s (T) & = \lim_{\bm{k} \rightarrow 0} \frac{T^2}{ B_0 (E_k, \bm{k};T)} = \lim_{ \bm{k} \rightarrow 0} \frac{T}{A (E_k, \bm{k};T) M_H} \ ,
\end{align}
where we express it with calligraphic font not to confuse it with $D_s$ mesons. 

In the left panel of this figure we show our result for this coefficient at low temperatures for $D$ mesons (solid red line) and for $\bar{B}$ mesons (dotted-dashed blue line). As expected, $\mathcal{ D}_s$ is lower for $\bar{B}$ mesons than for $D$ mesons, while the general trend is a monotonically decreasing function of temperature. As for the high temperature side, we plot two extractions from relativistic heavy-ion collisions at high energies using Bayesian analyses to estimate the temperature-behavior of this coefficient for the two flavors separately~\cite{Ke:2018tsh}. While a clear ordering cannot be settled in this extraction, a likely continuous matching can be observed around $T_c \simeq 155$ MeV. According to these results, the absolute minimum of the spatial diffusion coefficient might happen at the transition temperature $T_c$. On the right panel of Fig.~\ref{fig:transport-Ds} we plot our results at low temperatures together with results of a quasiparticle model for the quark-gluon plasma at high temperature, which can distinguish charm and bottom quarks~\cite{Das:2016llg}. An approximate matching is also seen in $T_c$ and the flavor-mass ordering of $\mathcal{D}_s$ is consistent in both sides of the transition. Close to $T_c$ a more refined model including a mixed phase with hadronization processes should be able to fill the gap.

\section{Conclusions}\label{sec:conclusions}
In this paper we have obtained the properties of mesons with open heavy-flavor at finite temperature using an effective field theory
based on chiral and HQSF symmetries within the imaginary-time formalism. The interaction of these pseudoscalar and vector open heavy-flavor ground-state mesons with light mesons ($\pi$, $K$, $\bar K$, $\eta$) is unitarized via a self-consistent coupled-channel Bethe-Salpeter approach at finite temperature.

With this methodology, we have obtained the self-energies and, hence, the corresponding spectral functions of open-charm ($D^{(*)}$, $D_s^{(*)}$) and open-bottom ($\bar B^{(*)}$, $\bar B_s^{(*)}$) ground states. On the one hand, we have determined that the values of the real part of self-energy over the heavy meson mass at a particular temperature are similar for $D_{(s)}$ and $D^*_{(s)}$, as well as for  $\bar B_{(s)}$ and  $\bar B^*_{(s)}$, as expected by HQSFS. On the other hand, the imaginary parts of the self-energies for the open heavy-flavor ground-state mesons becomes sizable with temperature due to the combination of the Landau and unitary cut effects at finite temperature. Therefore, the corresponding spectral functions shift towards lower energies and become wider with increasing temperature.  

From the behavior of the spectral functions, we have quantified the thermal dependence of the masses and the decay widths of the open heavy-flavor ground states. We have observed a generic downshift of the thermal masses with temperature, as large as of a few tens of MeV at $T = 150$ MeV in a pionic bath, while the decay widths increase with temperature up to values of some tens of MeV at $T = 150$ MeV. Compared to recent lattice QCD simulations for open-charm ground states~\cite{Aarts:2022krz}, a similar trend can be determined although a systematic shift is seen as a heavy non-physical pion mass is used in the lattice.

As a byproduct of the unitarization, we have also obtained the two-pole $D^*_0(2300)$ and $D_1(2430)$ as well as the $D^*_{s0}(2317)$ and $D_{s1}(2460)$  bound states (and the corresponding counterparts in the bottom sector) as dynamically generated by heavy-light meson scattering, and analyzed their behavior with temperature. The two-pole structures in the non-strange charm and bottom sectors gradually dilute with temperature with a smooth shift of their maxima, in spite of the difficulty to assess their evolution with temperature due to the closeness of the two-meson thresholds. As for the bound states, the $T=0$ delta-type states acquire non-zero width, and the shift and widening of the peak is comparable to that of the ground states.

And, finally, we have computed the transport coefficients for $D$ and $\bar B$ mesons propagating through an hadronic medium by means of an off-shell kinetic theory that is consistent with the effective field theory that describes the scattering processes of heavy mesons with light mesons at finite temperature. In particular,
we have obtained the drag force and the diffusion coefficients in momentum space, as well as the spatial diffusion one for both mesons. 

The diffusion coefficient of $\bar B$ turns out to be 2-3 times larger than that for the $D$ meson, whereas the drag coefficient for $\bar B$ becomes smaller (or comparable) to the $D$ meson one. This can be understood as the diffusion coefficient is proportional to the cross section (or imaginary part of the scattering amplitude) and this is larger for $\bar B$, whereas the drag force scales with the cross section but is also inversely proportional to the mass of the heavy meson. As for the spatial diffusion coefficients, the $D$ one is lower than that of the $\bar B$ meson as it is inversely proportional to the diffusion one in momentum space. Moreover, the spatial diffusion coefficients for $D$ and $\bar B$ mesons are monotonically decreasing functions of the temperature up to $T_c$, where a mininum might be present in order to match with the expected high-temperature behavior of the  coefficients in the QGP phase.

\section*{Author Contributions}

 All authors contributed equally to the paper. GM wrote the first draft of the manuscript. JT-R, LT, and AR wrote sections of the manuscript. All authors contributed to manuscript revision, read, and approved the submitted version.
 


\section*{Funding}
This research has been supported from the projects CEX2019-000918-M, CEX2020-001058-M (Unidades de Excelencia ``Mar\'{\i}a de Maeztu"), PID2019-110165GB-I00 and PID2020-118758GB-I00, financed by the Spanish MCIN/ AEI/10.13039/501100011033/, as well as by the EU STRONG-2020 project, under the program  H2020-INFRAIA-2018-1 grant agreement no. 824093. G.M. acknowledges support from the FPU17/04910 Doctoral Grant from the Spanish Ministerio de Universidades, and U.S. DOE Contract No. DE-AC05-06OR23177, under which Jefferson Science Associates, LLC, operates Jefferson Lab. L.T. and J.M.T.-R. acknowledge support from the DFG through projects no. 411563442 (Hot Heavy Mesons) and no. 315477589 - TRR 211 (Strong-interaction matter under extreme conditions). L.T. also acknowledges support from the Generalitat Valenciana under contract PROMETEO/2020/023.

\section*{Conflict of Interest Statement}

The authors declare that the research was conducted in the absence of any commercial or financial relationships that could be construed as a potential conflict of interest.

\bibliographystyle{apsrev4-2}
\bibliography{bib_revtex}
\end{document}